\documentclass[english,journal=jctcce,manuscript=review]{achemso}
\usepackage[T1]{fontenc}
\usepackage[latin9]{inputenc}
\usepackage{refstyle}
\usepackage{textcomp}
\usepackage{mathrsfs}
\usepackage{amsmath}
\usepackage{amssymb}
\usepackage{graphicx}
\PassOptionsToPackage{version=3}{mhchem}
\usepackage{mhchem}

\makeatletter

\title{Analytic Gradients of Approximate Coupled Cluster Methods with Quadruple
Excitations}
\author{Devin A. Matthews}
\affiliation{Southern Methodist University, Dallas, TX 75275, USA}
\email{damatthews@smu.edu}
\AtBeginDocument{\providecommand\subsecref[1]{\ref{subsec:#1}}}
\providecommand{\tabularnewline}{\\}
\RS@ifundefined{subsecref}
  {\newref{subsec}{name = \RSsectxt}}
  {}
\RS@ifundefined{thmref}
  {\def\RSthmtxt{theorem~}\newref{thm}{name = \RSthmtxt}}
  {}
\RS@ifundefined{lemref}
  {\def\RSlemtxt{lemma~}\newref{lem}{name = \RSlemtxt}}
  {}

\usepackage{mhchem}
\usepackage{simplewick}
\usepackage[binary-units=true]{siunitx}
\usepackage{pgfplots}
\usepgfplotslibrary{groupplots}

\allowdisplaybreaks

\@ifundefined{showcaptionsetup}{}{%
 \PassOptionsToPackage{caption=false}{subfig}}
\usepackage{subfig}
\makeatother

\usepackage{babel}
\begin{document}
\begin{abstract}
The analytic gradient theory for both iterative and non-iterative
coupled-cluster approximations that include connected quadruple excitations
is presented. These methods include, in particular, CCSDT(Q), which
is an analog of the well-known CCSD(T) method which starts from the
full CCSDT method rather than CCSD. The resulting methods are implemented
in the CFOUR program suite, and pilot applications are presented for
the equilibrium geometries and harmonic vibrational frequencies of
the simplest Criegee intermediate, $\ce{CH2OO}$, as well as to the
isomerization pathway between dimethylcarbene and propene. While all
methods are seen to approximate the full CCSDTQ results well for ``well-behaved''
systems, the more difficult case of the Criegee intermediate shows
that CCSDT(Q), as well as certain iterative approximations, display
problematic behavior.
\end{abstract}

\section{Introduction}

The inclusion of higher-order correlation effects has been recognized
as an important contribution to the calculation of accurate molecular
energies for more than a decade.\cite{martinChapterComputationalThermochemistry2005,fellerSurveyFactorsContributing2008,haunschildNewAccurateReference2012,dixonChapterOnePractical2012,petersonChemicalAccuracyInitio2012}
Such contributions are critical to the accurate evaluation of molecular
and reaction enthalpies,\cite{dixonHeatsFormationIonization2001,csaszarStandardEnthalpyFormation2003,fellerEnthalpyFormationN2H42017}
atomization energies,\cite{kartonAtomizationEnergiesCarbon2009,rudenCoupledclusterConnectedquadruplesCorrections2003}
barrier heights,\cite{wernerBarrierHeightH22008} and intermolecular
interaction energies,\cite{hopkinsInitioStudiesInteractions2004}
and are included in many standard thermochemical model chemistries
such as HEAT,\cite{tajtiHEATHighAccuracy2004,bombleHighaccuracyExtrapolatedInitio2006,hardingHighaccuracyExtrapolatedInitio2008,thorpeHighaccuracyExtrapolatedInitio2019}
W\emph{n},\cite{martinStandardMethodsBenchmark1999,boeseW3TheoryRobust2004,kartonW4TheoryComputational2006}
and ANL-\emph{n}.\cite{klippensteinInitioComputationsActive2017}
However, the effect of higher-order electron correlation on equilibrium
molecular geometries, vibrational frequencies, and other properties
is less well-understood. Including geometric effects beyond the CCSD(T)\cite{raghavachariFifthorderPerturbationComparison1989a}
or CCSDT\cite{nogaFullCCSDTModel1987a} level could be desirable in
a number of circumstances. For example, Morgan et al. calculated the
equilibrium geometry of formaldehyde including CCSDT(Q)\cite{bombleCoupledclusterMethodsIncluding2005}
and CCSDTQ\cite{kucharskiRecursiveIntermediateFactorization1991,oliphantCoupledClusterMethod1991,kucharskiCoupledClusterSingle1992}
higher-order correlation contributions, as well as anharmonic vibrational
frequencies including CCSDT(Q) corrections.\cite{morganGeometricEnergyDerivatives2018}
Puzzarini et al. investigated the effect of higher-order correlation
on computed rotational constants.\cite{puzzariniAccuracyRotationalConstants2008}
Heckert et al. computed the higher-order correlation contribution
to the geometry of several small molecules, and found rather large
changes in geometry especially for triply bonded species such as $\ce{N2}$,
$\ce{HCN}$, and $\ce{HCCH}$, as well as for $\ce{F2}$.\cite{heckertMolecularEquilibriumGeometries2005a,heckertBasissetExtrapolationTechniques2006}
Ruden et al. investigated the contribution of quadruple excitations
to the harmonic frequencies of several diatomics and found corrections
on going from CCSDT to CCSDTQ as large as 20 cm$^{-1}$.\cite{rudenCoupledclusterConnectedQuadruples2004}

The high computational expense of CCSDTQ naturally limits applicability
to small molecules. Approximate coupled cluster methods would ideally
extend the range of applicability, especially for non-iterative approximations
such as CCSDT(Q). However, for geometric derivatives and especially
for harmonic frequencies, the lack of analytic gradients further increases
the cost as finite-difference methods must be used. This additional
expense furthermore scales with the molecular size rather prohibitively
as the finite difference technique depends on the number of degrees
of freedom while analytic gradients do not.\cite{handyEvaluationAnalyticEnergy1984}
In order to efficiently study the effect of such higher-order effects
on molecular properties, we present here the derivation of analytic
gradients for a number of approximate coupled cluster methods that
include connected quadruple excitations: CCSDT(Q)\cite{bombleCoupledclusterMethodsIncluding2005}
(and its A and B variants\cite{kallayApproximateTreatmentHigher2008}),
CCSDTQ-1a, -1b, and -3,\cite{kallayApproximateTreatmentHigher2005}
and CC4.\cite{kallayApproximateTreatmentHigher2005} While analytic
gradients for general coupled cluster models, including CCSDT\cite{gaussAnalyticGradientsCoupledcluster2002}
and CCSDTQ,\cite{kallayAnalyticFirstDerivatives2003} have been available
for some time, analytic gradients for general approximate coupled
cluster methods\cite{kallayApproximateTreatmentHigher2005} have not
as yet been derived, except for the special cases of CCSDT-$n$,\cite{scuseriaAnalyticEvaluationEnergy1988,gaussAnalyticFirstSecond2000}
CC3,\cite{gaussAnalyticFirstSecond2000} and of course CCSD(T).\cite{scuseriaAnalyticEvaluationEnergy1991,wattsOpenshellAnalyticalEnergy1992}

\section{Theory}

The theory of analytic CCSDT(Q) gradients is developed by first reviewing
the basic theory of coupled cluster and its gradients.\cite{handyEvaluationAnalyticEnergy1984,scheinerAnalyticEvaluationEnergy1987,rendellEfficientFormulationImplementation1991,gaussCoupledClusterOpen1991}
Then, the derivation of the CCSDT(Q) energy is reviewed and considerations
for non-Hartree Fock references (such as ROHF and QRHF) are discussed.
These theories are then combined to derive explicit CCSDT(Q) gradient
expressions. Next, iterative approximations to the CCSDTQ energy are
reviewed, and finally the corresponding analytic gradient theories
are developed.

\subsection{Coupled Cluster Gradients}

The coupled cluster energy is conveniently written as a matrix element
of the coupled cluster transformed Hamiltonian, $\bar{H}=e^{-\hat{T}}\hat{H}e^{\hat{T}}=\left(\hat{H}e^{\hat{T}}\right)_{c}$,\cite{cizekCorrelationProblemAtomic1966,shavittManyBodyMethodsChemistry2009,helgakerMolecularElectronicStructureTheory2013}
\begin{equation}
E_{CC}=\langle0|\bar{H}|0\rangle\label{eq:cc-energy}
\end{equation}
where $\hat{H}$ is the Hamiltonian in the molecular orbital basis,
normal-ordered with respect to the reference single-particle wavefunction
$|0\rangle$, and the cluster operator $\hat{T}$ is an excitation
operator,
\begin{eqnarray}
\hat{H} & = & \hat{F}+\hat{V}=\sum_{pq}f_{q}^{p}\{p^{\dagger}q\}+\frac{1}{4}\sum_{pqrs}v_{rs}^{pq}\{p^{\dagger}q^{\dagger}sr\}\\
\hat{T} & = & \sum_{k=1}^{N}\hat{T}_{k}=\sum_{k=1}^{N}\frac{1}{(k!)^{2}}\sum_{\substack{a_{1}\ldots a_{k}\\
i_{1}\ldots i_{k}
}
}t_{i_{1}\ldots i_{k}}^{a_{1}\ldots a_{k}}a_{1}^{\dagger}\ldots a_{k}^{\dagger}i_{k}\ldots i_{1}
\end{eqnarray}
for occupied spin-orbitals $ij\ldots$, virtual (unoccupied) spin-orbitals
$ab\ldots$, and arbitrary spin-orbitals $pqrs$, and where $\{\ldots\}$
denotes normal ordering. The number of excitations $N$ included in
the cluster operator gives a hierarchy of coupled cluster methods,
CCSD ($N=2$) $\rightarrow$ CCSDT $\rightarrow$ CCSDTQ $\rightarrow\ldots\rightarrow$
Full Coupled Cluster (FCC), which is identical to the well-known and
exact Full Configuration Interaction (FCI) method.

However, when deriving the theory of coupled cluster gradients and
properties, it is more convenient to use the stationary coupled cluster
energy functional,\cite{scheinerAnalyticEvaluationEnergy1987}
\begin{align}
E_{CC} & =\langle0|(1+\hat{\Lambda})\bar{H}|0\rangle\\
\hat{\Lambda}=\sum_{k=1}^{N}\hat{\Lambda}_{k} & =\sum_{k=1}^{N}\frac{1}{(k!)^{2}}\sum_{\substack{a_{1}\ldots a_{k}\\
i_{1}\ldots i_{k}
}
}\lambda_{a_{1}\ldots a_{k}}^{i_{1}\ldots i_{k}}i_{1}^{\dagger}\ldots i_{k}^{\dagger}a_{k}\ldots a_{1}
\end{align}
Since the coupled cluster equations are satisfied,
\begin{equation}
0=\langle P|\bar{H}|0\rangle
\end{equation}
for all excited determinants $\langle P|=\langle S|+\langle D|+\langle T|+\langle Q|+\ldots$
(i.e. single, double, triple, quadruples excitations etc.) up to the
number of excitations included in the model, then the energy obtained
with this functional is trivially the same as in (\ref{eq:cc-energy}).
For CCSDT and CCSDT(Q), $\langle P|=\langle S|+\langle D|+\langle T|$.
However, both derivatives of the energy with respect to some parameter
$\chi$ and expectation values of an arbitrary property $\hat{O}$
can be succinctly written using this same functional,
\begin{eqnarray}
\frac{\partial E_{CC}}{\partial\chi}=E_{CC}^{\chi} & = & \langle0|(1+\hat{\Lambda})\bar{H}^{\chi}|0\rangle\label{eq:deriv}\\
\langle\hat{O}\rangle_{CC} & = & \langle0|(1+\hat{\Lambda})\bar{O}|0\rangle\label{eq:prop}
\end{eqnarray}
where $\bar{H}^{\chi}=\left(\hat{H}^{\chi}e^{\hat{T}}\right)_{c}=\left(\left(\frac{\partial\hat{H}}{\partial\chi}\right)e^{\hat{T}}\right)_{c}$
and $\bar{O}=\left(\hat{O}e^{\hat{T}}\right)_{c}$. These expressions
can be further generalized by constructing one- and two-particle density
matrices,
\begin{eqnarray}
E_{CC}^{\chi} & = & \sum_{pq}D_{q}^{p}\left(f_{q}^{p}\right)^{\chi}+\sum_{pqrs}\Gamma_{rs}^{pq}\left(v_{rs}^{pq}\right)^{\chi}\label{eq:cc-gradient-density}\\
\langle\hat{O}\rangle_{CC} & = & \sum_{pq}D_{q}^{p}o_{q}^{p}+\sum_{pqrs}\Gamma_{rs}^{pq}o_{rs}^{pq}\label{eq:exp-value-density}\\
D_{q}^{p} & = & \langle0|(1+\hat{\Lambda})\left(\{p^{\dagger}q\}\,e^{\hat{T}}\right)_{c}|0\rangle\\
\Gamma_{rs}^{pq} & = & \langle0|(1+\hat{\Lambda})\left(\{p^{\dagger}q^{\dagger}sr\}\,e^{\hat{T}}\right)_{c}|0\rangle
\end{eqnarray}
assuming that the operator $\hat{O}$ may have one-electron and/or
two-electron components.

The partial derivatives in (\ref{eq:cc-gradient-density}) and (\ref{eq:exp-value-density})
are derivatives of the molecular orbital basis quantities, and include
contributions from the response of the orbitals to the perturbation.\cite{riceAnalyticConfigurationInteraction1986,scheinerAnalyticEvaluationEnergy1987}
These contributions can be separated out and explicit dependence on
the perturbation removed through the use of the coupled perturbed
Hartree Fock (CPHF) $Z$-vector equations (similarly to how the $\hat{\Lambda}$
operator removes the dependence on the differentiated $\hat{T}$ amplitudes).\cite{handyEvaluationAnalyticEnergy1984}
The relationship of the orbital-response-corrected density to the
molecular orbital density is entirely independent of the source of
the density, and so is not specific to the particular flavor of coupled
cluster theory or even to coupled cluster at all. Similarly, as many
coupled cluster methods, including CCSDT(Q), require semi-canonical
orbitals (i.e. $f_{b}^{a}=\epsilon_{a}\delta_{ab}$ and $f_{j}^{i}=\epsilon_{i}\delta_{ij}$),
the modification of the density matrix to the use of perturbed canonical
orbitals is also independent of the source of the density matrix.\cite{scuseriaAnalyticEvaluationEnergy1991,wattsOpenshellAnalyticalEnergy1992}
This means that for methods which require (semi-)canonical orbitals,
only the diagonal elements of the virtual-virtual and occupied-occupied
one-electron density matrices need to be computed.

\subsection{The CCSDT(Q) Energy}

The derivation of the CCSDT(Q) energy is similar in many ways to the
derivation of the CCSD(T) energy. In the context of many-body perturbation
theory (MBPT), taking the lowest-order correction to the CCSDT energy,
and replacing the approximate $\hat{T}_{1}$, $\hat{T}_{2}$, and
$\hat{T}_{3}$ amplitudes with their converged CCSDT values gives
the CCSDT{[}Q{]} method of Kucharski and Bartlett,\cite{bartlettNoniterativeFifthorderTriple1990}
\begin{equation}
E_{[Q]}=\langle0|\hat{T}_{2}^{\dagger}\hat{V}\hat{R}_{4}\left(\hat{V}\left(\hat{T}_{3}+\frac{1}{2}\hat{T}_{2}^{2}\right)\right)_{c}|0\rangle
\end{equation}
where $\hat{R}_{4}$ is the resolvent operator in the quadruples space,
$\hat{R}_{4}=|Q\rangle\langle Q|\hat{F}|Q\rangle^{-1}\langle Q|$.
This approach is similar to that employed in the CCSD{[}T{]} (also
called CCSD+T(CCSD)) method. However, much as CCSD(T) significantly
improves on CCSD{[}T{]} through the inclusion of a single higher-order
term, CCSDT{[}Q{]} can be improved upon.

Following the analogue of the derivation of CCSD(T) by Stanton\cite{stantonWhyCCSDWorks1997}
(as done by Bomble et al.\cite{bombleCoupledclusterMethodsIncluding2005}
and Kállay et al.\cite{kallayApproximateTreatmentHigher2005}), the
exact energy $E$ can be written by applying the Löwdin partitioning
approach and then expanded in perturbation orders along with the transformed
Hamiltonian,
\begin{eqnarray}
E & = & \langle0|(1+\hat{\Lambda}_{CCSDT})\bar{H}_{CCSDT}|\tilde{P}\rangle\langle\tilde{P}|\left(\bar{H}_{CCSDT}-E\right)^{-1}|\tilde{P}\rangle\langle\tilde{P}|\bar{H}_{CCSDT}|0\rangle\\
E & = & E^{[0]}+E^{[1]}+E^{[2]}+\ldots\\
\bar{H}_{CCSDT} & = & \bar{H}_{CCSDT}^{[0]}+\bar{H}_{CCSDT}^{[1]}+\bar{H}_{CCSDT}^{[2]}+\ldots
\end{eqnarray}
where $\hat{\Lambda}_{CCSDT}=\hat{\Lambda}_{1}+\hat{\Lambda}_{2}+\hat{\Lambda}_{3}$,
$\bar{H}_{CCSDT}=(\hat{H}e^{\hat{T}_{1}+\hat{T}_{2}+\hat{T}_{3}})_{c}$,
and $|\tilde{P}\rangle=|1-P\rangle=|Q\rangle+\ldots$ is the complement
space. Perturbation orders are assigned according to the usual Møller-Plesset
partitioning. Taking the lowest (fourth)-order contribution gives
the CCSDT(Q)$_{\Lambda}$ (also called $\Lambda$CCSDT(Q)) energy,
\begin{eqnarray}
E_{(Q)_{\Lambda}} & = & \langle0|(1+\hat{\Lambda}_{CCSDT})\bar{H}_{CCSDT}^{[1]}|Q\rangle\langle Q|\left(\bar{H}_{CCSDT}^{[0]}-E^{[0]}\right)^{-1}|Q\rangle\langle Q|\bar{H}_{CCSDT}^{[3]}|0\rangle\\
 & = & \langle0|\left(\hat{\Lambda}_{2}+\hat{\Lambda}_{3}\right)\hat{V}\hat{T}_{4}^{[3]}|0\rangle\\
\hat{T}_{4}^{[3]} & = & \hat{R}_{4}\left(\hat{V}\left(\hat{T}_{3}+\frac{1}{2}\hat{T}_{2}^{2}\right)\right)_{c}
\end{eqnarray}

For canonical Hartree Fock references (or other references where $f_{i}^{a}=f_{a}^{i}=0$),
the converged $\hat{\Lambda}_{2}$ and $\hat{\Lambda}_{3}$ amplitudes
are the same as $\hat{T}_{2}$ and $\hat{T}_{3}$ to lowest order.
Thus, as in CCSD(T), the $\hat{\Lambda}$ amplitudes may be approximated
by $\hat{T}^{\dagger}$, giving the CCSDT(Q) method,
\begin{equation}
E_{(Q)}=\langle0|\left(\hat{T}_{2}^{\dagger}+\hat{T}_{3}^{\dagger}\right)\hat{V}\hat{T}_{4}^{[3]}|0\rangle
\end{equation}
For non-Hartree Fock references such as ROHF and QRHF, though, Kállay
notes that $\hat{T}_{3}$ and $\hat{\Lambda}_{3}$ are no longer identical
at lowest order.\cite{kallayApproximateTreatmentHigher2008} By incorporating
the additional lowest-order (disconnected) contributions and replacing
approximate $\hat{T}_{1}$ and $\hat{T}_{2}$ amplitudes with their
converged counterparts at one of two stages, two additional methods,
CCSDT(Q)/A and CCSDT(Q)/B may be derived,
\begin{eqnarray}
E_{(Q)/A} & = & \langle0|\left(\hat{T}_{2}^{\dagger}+\hat{T}_{3}^{\dagger}+\hat{T}_{1}^{\dagger}\hat{T}_{2}^{\dagger}\right)\hat{H}\hat{T}_{4}^{[3]}|0\rangle\\
E_{(Q)/B} & = & \langle0|\left(\hat{T}_{2}^{\dagger}+\hat{T}_{3}^{\dagger}+\left(\hat{T}_{1}^{\dagger}\hat{V}+\hat{T}_{2}^{\dagger}\hat{F}\right)\hat{R}_{3}\right)\hat{H}\hat{T}_{4}^{[3]}|0\rangle\nonumber \\
 & = & \langle0|\left(\hat{T}_{2}^{\dagger}+\hat{T}_{3}^{\dagger}+\hat{T}_{3}^{B}\right)\hat{H}\hat{T}_{4}^{[3]}|0\rangle
\end{eqnarray}

One difficulty with these methods, though, is that when applied to
a canonical reference, they do not reduce to the ``normal'' CCSDT(Q)
energy. While the additional contributions to $\hat{\Lambda}_{3}$
in the non-HF case are technically at the same order as the portion
in common with $\hat{T}_{3}$, these terms are entirely disconnected,
which tends to produce a numerically less significant contribution.
On the other hand, the balance between connected and disconnected
terms is sometimes necessary to avoid excessive basis set dependency
and other problems, as in the case of CCSDT{[}Q{]} vs. CCSDT(Q). The
gradients of the normal HF CCSDT(Q) method as well as the two non-HF
variants will be examined on an equal footing.

\subsection{CCSDT(Q) Gradients}

The gradient of the CCSDT(Q) energy can be simply obtained by differentiating
the energy expression:
\begin{eqnarray}
E_{(Q)}^{\chi} & = & \langle0|\left(\hat{T}_{2}^{\chi\dagger}+\hat{T}_{3}^{\chi\dagger}\right)\hat{V}\hat{T}_{4}^{[3]}|0\rangle+\langle0|\left(\hat{T}_{2}^{\dagger}+\hat{T}_{3}^{\dagger}\right)\hat{V}^{\chi}\hat{T}_{4}^{[3]}|0\rangle-\langle0|\hat{T}_{4}^{\prime}\hat{F}^{\chi}\hat{T}_{4}^{[3]}|0\rangle\nonumber \\
 &  & +\langle0|\hat{T}_{4}^{\prime}\left(\hat{V}^{\chi}\left(\hat{T}_{3}+\frac{1}{2}\hat{T}_{2}^{2}\right)\right)_{c}|0\rangle+\langle0|\hat{T}_{4}^{\prime}\left(\hat{V}\left(\hat{T}_{3}^{\chi}+\hat{T}_{2}\hat{T}_{2}^{\chi}\right)\right)_{c}|0\rangle\\
E_{(Q)/A}^{\chi} & = & \langle0|\left(\hat{T}_{2}^{\chi\dagger}+\hat{T}_{3}^{\chi\dagger}+\hat{T}_{1}^{\chi\dagger}\hat{T}_{2}^{\dagger}+\hat{T}_{1}\hat{T}_{2}^{\chi\dagger}\right)\hat{H}\hat{T}_{4}^{[3]}|0\rangle+\langle0|\left(\hat{T}_{2}^{\dagger}+\hat{T}_{3}^{\dagger}+\hat{T}_{1}^{\dagger}\hat{T}_{2}^{\dagger}\right)\hat{H}^{\chi}\hat{T}_{4}^{[3]}|0\rangle\nonumber \\
 &  & -\langle0|\hat{T}_{4}^{A}\hat{F}^{\chi}\hat{T}_{4}^{[3]}|0\rangle+\langle0|\hat{T}_{4}^{A}\left(\hat{V}^{\chi}\left(\hat{T}_{3}+\frac{1}{2}\hat{T}_{2}^{2}\right)\right)_{c}|0\rangle\nonumber \\
 &  & +\langle0|\hat{T}_{4}^{A}\left(\hat{V}\left(\hat{T}_{3}^{\chi}+\hat{T}_{2}\hat{T}_{2}^{\chi}\right)\right)_{c}|0\rangle\\
E_{(Q)/B}^{\chi} & = & \langle0|\left(\hat{T}_{2}^{\chi\dagger}+\hat{T}_{3}^{\chi\dagger}+\left(\hat{T}_{1}^{\chi\dagger}\hat{V}+\hat{T}_{2}^{\chi\dagger}\hat{F}\right)\hat{R}_{3}\right)\hat{H}\hat{T}_{4}^{[3]}|0\rangle\nonumber \\
 &  & +\langle0|\left(\hat{T}_{1}^{\dagger}\hat{V}^{\chi}+\hat{T}_{2}^{\dagger}\hat{F}^{\chi}\right)\hat{R}_{3}\hat{H}\hat{T}_{4}^{[3]}|0\rangle-\langle0|\hat{T}_{3}^{B}\hat{F}^{\chi}\hat{R}_{3}\hat{H}\hat{T}_{4}^{[3]}|0\rangle\nonumber \\
 &  & +\langle0|\left(\hat{T}_{2}^{\dagger}+\hat{T}_{3}^{\dagger}+\hat{T}_{3}^{B}\right)\hat{H}^{\chi}\hat{T}_{4}^{[3]}|0\rangle-\langle0|\hat{T}_{4}^{B}\hat{F}^{\chi}\hat{T}_{4}^{[3]}|0\rangle\nonumber \\
 &  & +\langle0|\hat{T}_{4}^{B}\left(\hat{V}^{\chi}\left(\hat{T}_{3}+\frac{1}{2}\hat{T}_{2}^{2}\right)\right)_{c}|0\rangle+\langle0|\hat{T}_{4}^{B}\hat{R}_{4}\left(\hat{V}\left(\hat{T}_{3}^{\chi}+\hat{T}_{2}\hat{T}_{2}^{\chi}\right)\right)_{c}|0\rangle
\end{eqnarray}
where we have defined convenient ``left-hand'' $\hat{T}_{4}$ intermediates,
\begin{align}
\hat{T}_{4}^{\prime} & =\left(\hat{T}_{2}^{\dagger}+\hat{T}_{3}^{\dagger}\right)\hat{V}\hat{R}_{4}\\
\hat{T}_{4}^{A} & =\left(\hat{T}_{2}^{\dagger}+\hat{T}_{3}^{\dagger}+\hat{T}_{1}^{\dagger}\hat{T}_{2}^{\dagger}\right)\hat{H}\hat{R}_{4}\\
\hat{T}_{4}^{B} & =\left(\hat{T}_{2}^{\dagger}+\hat{T}_{3}^{\dagger}+\hat{T}_{3}^{B}\right)\hat{H}\hat{R}_{4}
\end{align}
and used the identity $\hat{R}_{n}^{\chi}=-\hat{R}_{n}\hat{F}^{\chi}\hat{R}_{n}$.
We can separate each of these expressions into two parts: one which
depends on derivatives of the Hamiltonian multiplied by CCSDT(Q) density
matrix elements, and one which depends on derivatives of the coupled
cluster amplitudes:
\begin{eqnarray}
E_{(Q)}^{\chi} & = & \langle0|D_{(Q)}^{\prime}\hat{F}^{\chi}|0\rangle+\langle0|\Gamma_{(Q)}^{\prime}\hat{V}^{\chi}|0\rangle+\langle0|\hat{S}_{2}\hat{T}_{2}^{\chi}|0\rangle+\langle0|\hat{S}_{3}\hat{T}_{3}^{\chi}|0\rangle\\
E_{(Q)/A,B}^{\chi} & = & \langle0|D_{(Q)/A,B}^{\prime}\hat{F}^{\chi}|0\rangle+\langle0|\Gamma_{(Q)/A,B}^{\prime}\hat{V}^{\chi}|0\rangle+\langle0|\hat{S}_{1}^{A/B}\hat{T}_{1}^{\chi}|0\rangle\nonumber \\
 &  & +\langle0|\hat{S}_{2}^{A/B}\hat{T}_{2}^{\chi}|0\rangle+\langle0|\hat{S}_{3}^{A/B}\hat{T}_{3}^{\chi}|0\rangle
\end{eqnarray}
given in terms of the density matrices,
\begin{eqnarray}
\left(D_{(Q)}^{\prime}\right)_{q}^{p} & = & -\delta_{pq}\langle0|\hat{T}_{4}^{\prime}\{p^{\dagger}q\}\,\hat{T}_{4}^{[3]}|0\rangle\\
\left(D_{(Q)/A}^{\prime}\right)_{q}^{p} & = & \Delta_{pq}\langle0|\left(\hat{T}_{3}^{\dagger}+\hat{T}_{1}^{\dagger}\hat{T}_{2}^{\dagger}-\hat{T}_{4}^{A}\right)\{p^{\dagger}q\}\,\hat{T}_{4}^{[3]}|0\rangle\\
\left(D_{(Q)/B}^{\prime}\right)_{q}^{p} & = & \Delta_{pq}\left\{ \langle0|\left(\hat{T}_{3}^{\dagger}+\hat{T}_{3}^{B}-\hat{T}_{4}^{B}\right)\{p^{\dagger}q\}\,\hat{T}_{4}^{[3]}|0\rangle\right.\nonumber \\
 &  & +\langle\left.0|\left(\hat{T}_{2}^{\dagger}-\hat{T}_{3}^{B}\right)\{p^{\dagger}q\}\,\hat{R}_{3}\hat{H}\hat{T}_{4}^{[3]}|0\rangle\right\} \\
\left(\Gamma_{(Q)}^{\prime}\right)_{rs}^{pq} & = & \langle0|\left(\hat{T}_{2}^{\dagger}+\hat{T}_{3}^{\dagger}\right)\{p^{\dagger}q^{+}sr\}\,\hat{T}_{4}^{[3]}|0\rangle+\langle0|\hat{T}_{4}^{\prime}\left(\{p^{\dagger}q^{\dagger}sr\}\left(\hat{T}_{3}+\frac{1}{2}\hat{T}_{2}^{2}\right)\right)_{c}|0\rangle\label{eq:(q)-2pdm}\\
\left(\Gamma_{(Q)/A}^{\prime}\right)_{rs}^{pq} & = & \langle0|\left(\hat{T}_{2}^{\dagger}+\hat{T}_{3}^{\dagger}+\hat{T}_{1}^{\dagger}\hat{T}_{2}^{\dagger}\right)\{p^{\dagger}q^{\dagger}sr\}\,\hat{T}_{4}^{[3]}|0\rangle\nonumber \\
 &  & +\langle0|\hat{T}_{4}^{A}\left(\{p^{\dagger}q^{\dagger}sr\}\left(\hat{T}_{3}+\frac{1}{2}\hat{T}_{2}^{2}\right)\right)_{c}|0\rangle\\
\left(\Gamma_{(Q)/B}^{\prime}\right)_{rs}^{pq} & = & \langle0|\left(\hat{T}_{2}^{\dagger}+\hat{T}_{3}^{\dagger}+\hat{T}_{3}^{B}\right)\{p^{\dagger}q^{\dagger}sr\}\,\hat{T}_{4}^{[3]}|0\rangle+\langle0|\hat{T}_{1}^{\dagger}\{p^{\dagger}q^{\dagger}sr\}\,\hat{R}_{3}\hat{H}\hat{T}_{4}^{[3]}|0\rangle\nonumber \\
 &  & +\langle0|\hat{T}_{4}^{B}\left(\{p^{\dagger}q^{\dagger}sr\}\left(\hat{T}_{3}+\frac{1}{2}\hat{T}_{2}^{2}\right)\right)_{c}|0\rangle
\end{eqnarray}
where $\Delta_{pq}=\delta_{pq}+\delta_{pa}\delta_{qi}+\delta_{pi}\delta_{qa}$,
and intermediates,
\begin{eqnarray}
\hat{S}_{1}^{A} & = & \hat{T}_{4}^{[3]\dagger}\hat{H}\hat{T}_{2}|S\rangle\langle S|\\
\hat{S}_{1}^{B} & = & \hat{T}_{4}^{[3]\dagger}\hat{H}\hat{R}_{3}\hat{V}|S\rangle\langle S|\\
\hat{S}_{2} & = & \contraction[0.5ex]{\left(\hat{T}_{4}^{[3]\dagger}\hat{V}+\hat{T}_{4}^{\prime}\right.\left(\right.}{\hat{V}}{\left.\left.\!\!\hat{T}_{2}\right)_{c}\right)|}{D}\left(\hat{T}_{4}^{[3]\dagger}\hat{V}+\hat{T}_{4}^{\prime}\right.\left(\hat{V}\right.\left.\left.\!\!\hat{T}_{2}\right)_{c}\right)|D\rangle\langle D|\\
\hat{S}_{2}^{A} & = & \contraction[0.5ex]{\left(\hat{T}_{4}^{[3]\dagger}\left(\hat{V}+\hat{H}\hat{T}_{1}\right)+\hat{T}_{4}^{A}\right.\left(\right.}{\hat{V}}{\left.\left.\!\!\hat{T}_{2}\right)_{c}\right)|}{D}\left(\hat{T}_{4}^{[3]\dagger}\left(\hat{V}+\hat{H}\hat{T}_{1}\right)+\hat{T}_{4}^{A}\right.\left(\hat{V}\right.\left.\left.\!\!\hat{T}_{2}\right)_{c}\right)|D\rangle\langle D|\\
\hat{S}_{2}^{B} & = & \contraction[0.5ex]{\left(\hat{T}_{4}^{[3]\dagger}\left(\hat{V}+\hat{H}\hat{R}_{3}\hat{F}\right)+\hat{T}_{4}^{B}\right.\left(\right.}{\hat{V}}{\left.\left.\!\!\hat{T}_{2}\right)_{c}\right)|}{D}\left(\hat{T}_{4}^{[3]\dagger}\left(\hat{V}+\hat{H}\hat{R}_{3}\hat{F}\right)+\hat{T}_{4}^{B}\right.\left(\hat{V}\right.\left.\left.\!\!\hat{T}_{2}\right)_{c}\right)|D\rangle\langle D|\\
\hat{S}_{3} & = & \left(\hat{T}_{4}^{[3]\dagger}+\hat{T}_{4}^{\prime}\right)\hat{V}|T\rangle\langle T|\\
\hat{S}_{3}^{A,B} & = & \left(\hat{T}_{4}^{[3]\dagger}\hat{H}+\hat{T}_{4}^{A,B}\hat{V}\right)|T\rangle\langle T|
\end{eqnarray}
where the contraction line indicates that at least one creation/annihilation
operator in $\hat{V}$ must remain uncontracted (i.e. $\hat{V}$ must
have at least one external line in the diagrammatic representation).
The factor $\Delta_{pq}$ indicates that the density matrix elements
only need be computed for the occupied-virtual and virtual-occupied
blocks and for the diagonal of the occupied-occupied and virtual-virtual
blocks. Note that the CCSDT(Q) one-particle density matrix does not
have an occupied-virtual contribution, and is only valid for canonical
(HF) references.

At this point the equations for the CCSDT(Q) and CCSDT(Q)/A,B gradients
may be combined in a generic expression,
\begin{equation}
E_{(Q)/X}^{\chi}=\langle0|D_{(Q)/X}^{\prime}\hat{F}^{\chi}|0\rangle+\langle0|\Gamma_{(Q)/X}^{\prime}\hat{V}^{\chi}|0\rangle+\langle0|\hat{S}^{X}\hat{T}^{\chi}|0\rangle
\end{equation}
where $\hat{S}^{X}=\hat{S}_{1}^{X}+\hat{S}_{2}^{X}+\hat{S}_{3}^{X}$
and either $X=A,B$ or is ``empty'', indicating the CCSDT(Q) gradient
and intermediates (with $\hat{S}_{1}=0$). Now, the definition of
the derivative coupled cluster amplitudes (obtained by differentiating
the coupled cluster amplitude equations) may be inserted to transform
dependence on the derivative amplitudes to dependence on the derivative
Hamiltonian,
\begin{align}
E_{(Q)/X}^{\chi} & =\langle0|D_{(Q)/X}^{\prime}\hat{F}^{\chi}|0\rangle+\langle0|\Gamma_{(Q)/X}^{\prime}\hat{V}^{\chi}|0\rangle-\langle0|\hat{S}^{X}\left(\bar{H}-E_{CCSDT}\right)^{-1}|P\rangle\langle P|\bar{H}^{\chi}|0\rangle\nonumber \\
 & =\langle0|D_{(Q)/X}^{\prime}\hat{F}^{\chi}|0\rangle+\langle0|\Gamma_{(Q)/X}^{\prime}\hat{V}^{\chi}|0\rangle+\langle0|\tilde{\Lambda}^{X}\bar{H}^{\chi}|0\rangle
\end{align}

Lastly, the solution of the $\tilde{\Lambda}^{X}$ equations can be
combined with the solution of the CCSDT $\hat{\Lambda}$ equations
by solving for their linear combination,
\begin{equation}
\langle0|\left(\hat{\Lambda}+\tilde{\Lambda}^{X}\right)\left(\bar{H}-E_{CCSDT}\right)|P\rangle+\langle0|\left(\bar{H}+\hat{S}^{X}\right)|P\rangle=0
\end{equation}
and the combined density matrices constructed which give the gradient
of the total CCSDT(Q)/X energy (total correlation energy; the gradient
of the reference energy is computed and added in the standard way),
\begin{align}
E_{CCSDT(Q)/X}^{\chi} & =\langle0|D_{CCSDT(Q)/X}\hat{F}^{\chi}|0\rangle+\langle0|\Gamma_{CCSDT(Q)/X}\hat{V}^{\chi}|0\rangle\\
\left(D_{CCSDT(Q)/X}\right)_{q}^{p} & =\left(D_{(Q)/X}^{\prime}\right)_{q}^{p}+\langle0|\left(1+\hat{\Lambda}+\tilde{\Lambda}^{X}\right)\left(\{p^{\dagger}q\}\,e^{\hat{T}}\right)_{c}|0\rangle\label{eq:1pdm}\\
\left(\Gamma_{CCSDT(Q)/X}\right)_{rs}^{pq} & =\left(\Gamma_{(Q)/X}^{\prime}\right)_{rs}^{pq}+\langle0|\left(1+\hat{\Lambda}+\tilde{\Lambda}^{X}\right)\left(\{p^{\dagger}q^{\dagger}sr\}\,e^{\hat{T}}\right)_{c}|0\rangle\label{eq:2pdm}
\end{align}
These density matrices may then be processed in the usual way to compute
energy gradients and properties as desired.

\subsection{Iterative Approximations}

The CCSDT(Q) energy is a non-iterative correction that is applied
after the CCSDT equations have converged. Another approach to approximating
the CCSDTQ energy may be obtained by deleting some terms from the
CCSDTQ equations, but maintaining the iterative structure of the problem.
In particular, we wish to delete at least all $\hat{T}_{4}\rightarrow\hat{T}_{4}$
terms except those coming from $\hat{H}^{[0]}\equiv\sum_{p}f_{p}^{p}\{p^{\dagger}p\}$.
This leads to an equation for the $\hat{T}_{4}$ amplitudes of the
form $0=\langle Q|\tilde{H}+\hat{H}^{[0]}\hat{T}_{4}|0\rangle$ or
equivalently $\hat{T}_{4}=\hat{R}_{4}\langle Q|\tilde{H}|0\rangle$,
where $\tilde{H}$ is an effective operator discussed below. Because
the $\hat{T}_{4}$ equations can be solved exactly (given $\hat{T}_{1}$--$\hat{T}_{3}$
amplitudes which may solve their own equations only approximately),
the $\hat{T}_{4}$ amplitudes will only appear as ``intermediates''---directly
constructed from $\hat{T}_{1}$, $\hat{T}_{2}$, and $\hat{T}_{3}$
and then immediately consumed in the remaining amplitudes equations.
This structure both eliminates the costly $\mathscr{O}(n^{10})$ steps
of the CCSDTQ equations and allows for reduced storage of $\hat{T}_{4}$
since it may be immediately calculated, used, and discarded.

Depending on the additional terms that are deleted (guided by a mixture
of perturbation theory and pragmatism), the following approximations
may derived,\cite{kallayApproximateTreatmentHigher2005}

\bigskip{}

CCSDTQ-1a ($\equiv\text{CCSDTQ-1}$):
\begin{align}
\hat{T}_{4}^{CCSDTQ-1a} & =\hat{R}_{4}\left(\hat{V}\left(\hat{T}_{3}+\tfrac{1}{2}\hat{T}_{2}^{2}\right)\right)_{c}=\hat{T}_{4}^{[3]}\\
0 & =\langle S|\bar{H}_{CCSDT}|0\rangle\\
0 & =\langle D|\bar{H}_{CCSDT}+\hat{V}\hat{T}_{4}^{CCSDTQ-1a}]|0\rangle\\
0 & =\langle T|\bar{H}_{CCSDT}|0\rangle
\end{align}

CCSDTQ-1b:
\begin{align}
\hat{T}_{4}^{CCSDTQ-1b} & =\hat{R}_{4}\left(\hat{V}\left(\hat{T}_{3}+\tfrac{1}{2}\hat{T}_{2}^{2}\right)\right)_{c}\\
0 & =\langle S|\bar{H}_{CCSDT}|0\rangle\\
0 & =\langle D|\bar{H}_{CCSDT}+\hat{V}\hat{T}_{4}^{CCSDTQ-1b}]|0\rangle\\
0 & =\langle T|\bar{H}_{CCSDT}+\left(\left(\hat{H}+\hat{V}\hat{T}_{1}\right)\hat{T}_{4}^{CCSDTQ-1b}\right)_{c}|0\rangle
\end{align}

CCSDTQ-3:
\begin{align}
\hat{T}_{4}^{CCSDTQ-3} & =\hat{R}_{4}\left(\hat{H}e^{\hat{T}_{1}+\hat{T}_{2}}\left(1+\hat{T}_{3}\right)+\left(\hat{P}_{T}\hat{V}\hat{T}_{3}\hat{P}_{D}\right)\hat{T}_{3}\right)_{c}\\
0 & =\langle S|\bar{H}_{CCSDT}|0\rangle\\
0 & =\langle D|\bar{H}_{CCSDT}+\hat{V}\hat{T}_{4}^{CCSDTQ-3}]|0\rangle\\
0 & =\langle T|\bar{H}_{CCSDT}+\left(\left(\hat{H}+\hat{V}\hat{T}_{1}\right)\hat{T}_{4}^{CCSDTQ-3}\right)_{c}|0\rangle
\end{align}

CC4:
\begin{align}
\hat{T}_{4}^{CC4} & =\hat{R}_{4}\left(\hat{V}^{\prime}\left(\hat{T}_{3}+\tfrac{1}{2}\hat{T}_{2}^{2}\right)\right)\\
0 & =\langle S|\bar{H}_{CCSDT}|0\rangle\\
0 & =\langle D|\bar{H}_{CCSDT}+\hat{V}\hat{T}_{4}^{CC4}]|0\rangle\\
0 & =\langle T|\bar{H}_{CCSDT}+\left(\left(\hat{H}+\hat{V}\hat{T}_{1}\right)\hat{T}_{4}^{CC4}\right)_{c}|0\rangle
\end{align}

\noindent where $\hat{P}_{X}=|X\rangle\langle X|$ is a projection
operator onto the given excitation manifold and $\hat{V}^{\prime}=(\hat{V}e^{\hat{T}_{1}})_{c}$
is the $\hat{T}_{1}$-transformed two-electron potential. Note that
$\hat{T}_{4}^{CCSDTQ-3}$ is almost equal to $\hat{R}_{4}\langle Q|\bar{H}_{CCSDT}|0\rangle$,
but that it differs by the two terms $t_{ijkl}^{abcd}\gets\frac{1}{4}v_{ef}^{mn}t_{ijk}^{efc}t_{mnl}^{abd}+\frac{1}{2}v_{ef}^{mn}t_{mij}^{eab}t_{nkl}^{fcd}$
which are excluded by $(\hat{P}_{T}\hat{V}\hat{T}_{3}\hat{P}_{D})\hat{T}_{3}$
(since the projection operators and connectivity condition require
that $\hat{V}$ be connected to the inner $\hat{T}_{3}$ vertex by
exactly three indices). These terms are specifically deleted as they
would require $\mathscr{O}(n^{10})$ computation.

\subsection{Gradients in the Iterative Approximation}

The derivation of the gradient of the energy in the iterative approximation
closely follows the derivation of gradients for canonical CC methods.
However, since the cluster equations have been modified to delete
specific terms, we express the energy and its derivatives using an
effective transformed Hamiltonian $\tilde{H}$ (also called the Jacobian),
\begin{align}
E_{X} & =\langle0|(1+\hat{\Lambda})\tilde{H}_{X}|0\rangle\label{eq:approx-ener}\\
E_{X}^{\chi} & =\langle0|(1+\hat{\Lambda})\tilde{H}_{X}^{\chi}|0\rangle\label{eq:approx-grad}
\end{align}
where $X$ is one of the approximate methods above. The effective
transformed Hamiltonian may be specified in block form for each of
the approximate methods,
\begin{equation}
\tilde{H}_{X}=\begin{array}{cc}
0\quad\;\;\;S\quad\;\;\;D\quad\;\;\;T\quad\;\;\;Q\\
\left(\begin{array}{ccccc}
E_{X} & \bar{H}_{0S} & \hat{V} & 0 & 0\\
\tilde{H}_{S0} & \bar{H}_{SS} & \bar{H}_{SD} & \hat{V} & 0\\
\tilde{H}_{D0} & \bar{H}_{DS} & \bar{H}_{DD} & \bar{H}_{DT} & \hat{V}\\
\tilde{H}_{T0} & \bar{H}_{TS} & \bar{H}_{TD} & \bar{H}_{TT} & \tilde{H}_{TQ}\\
\tilde{H}_{Q0} & \tilde{H}_{QS} & \tilde{H}_{QD} & \tilde{H}_{QT} & \hat{H}^{[0]}
\end{array}\right) & \begin{array}{c}
0\\
S\\
D\\
T\\
Q
\end{array}
\end{array}
\end{equation}
The $\bar{H}$ blocks are identical to the corresponding blocks of
$\bar{H}_{CCSDT}$. $\tilde{H}_{TQ}$ is zero for CCSDTQ-1a and equal
to $\hat{H}+[\hat{V},\hat{T}_{1}]$ for all other methods. The remaining
blocks may be derived for each method from the corresponding equation
for $\hat{T}_{4}$, such that $\hat{T}_{4}=\hat{R}_{4}\tilde{H}_{Q0}$,
$\frac{\partial\hat{T}_{4}}{\partial\hat{T}_{1}}=\hat{R}_{4}\tilde{H}_{QS}$,
and similarly for $\tilde{H}_{QD}$ and $\tilde{H}_{QT}$, and from
the modified amplitude equations, $\tilde{H}_{S0}=\tilde{H}_{D0}=\tilde{H}_{T0}=0$.

The stationarity conditions of the energy functional ((\ref{eq:approx-ener})
and (\ref{eq:approx-grad})) along with the definition of $\tilde{H}$
determine the equations for $\hat{\Lambda}$,

\bigskip{}

CCSDTQ-1a:
\begin{align}
\hat{\Lambda}_{4}^{CCSDTQ-1a} & =\hat{\Lambda}_{2}\hat{V}\hat{R}_{4}=\hat{\Lambda}_{4}^{[2]}\\
0 & =\langle0|(1+\hat{\Lambda})\contraction[0.5ex]{}{\bar{H}}{|}{S}\bar{H}|S\rangle_{CCSDT}\\
0 & =\langle0|(1+\hat{\Lambda})\contraction[0.5ex]{}{\bar{H}}{|}{D}\bar{H}|D\rangle_{CCSDT}+\langle0|\hat{\Lambda}_{4}^{CCSDTQ-1a}\contraction[0.5ex]{[}{\hat{V}}{,\hat{T}_{2}]|}{D}[\hat{V},\hat{T}_{2}]|D\rangle\\
0 & =\langle0|(1+\hat{\Lambda})\contraction[0.5ex]{}{\bar{H}}{|}{T}\bar{H}|T\rangle_{CCSDT}\mathrel{+}\langle0|\hat{\Lambda}_{4}^{CCSDTQ-1a}\hat{V}|T\rangle
\end{align}

CCSDTQ-1b:
\begin{align}
\hat{\Lambda}_{4}^{CCSDTQ-1b} & =\left(\hat{\Lambda}_{2}\hat{V}+\hat{\Lambda}_{3}\left(\hat{H}+[\hat{V},\hat{T}_{1}]\right)\right)\hat{R}_{4}\\
0 & =\langle0|(1+\hat{\Lambda})\contraction[0.5ex]{}{\bar{H}}{|}{S}\bar{H}|S\rangle_{CCSDT}+\langle0|\hat{\Lambda}_{3}\contraction[0.5ex]{}{\hat{V}}{\hat{T}_{4}^{CCSDTQ-1b}|}{S}\hat{V}\hat{T}_{4}^{CCSDTQ-1b}|S\rangle\\
0 & =\langle0|(1+\hat{\Lambda})\contraction[0.5ex]{}{\bar{H}}{|}{D}\bar{H}|D\rangle_{CCSDT}+\langle0|\hat{\Lambda}_{4}^{CCSDTQ-1b}\contraction[0.5ex]{[}{\hat{V}}{,\hat{T}_{2}]|}{D}[\hat{V},\hat{T}_{2}]|D\rangle\\
0 & =\langle0|(1+\hat{\Lambda})\contraction[0.5ex]{}{\bar{H}}{|}{T}\bar{H}|T\rangle_{CCSDT}\mathrel{+}\langle0|\hat{\Lambda}_{4}^{CCSDTQ-1b}\hat{V}|T\rangle
\end{align}

CCSDTQ-3:
\begin{align}
\hat{\Lambda}_{4}^{CCSDTQ-3} & =\left(\hat{\Lambda}_{2}\hat{V}+\hat{\Lambda}_{3}\left(\hat{H}+[\hat{V},\hat{T}_{1}]\right)\right)\hat{R}_{4}\\
0 & =\langle0|(1+\hat{\Lambda})\contraction[0.5ex]{}{\bar{H}}{|}{S}\bar{H}|S\rangle_{CCSDT}+\langle0|\hat{\Lambda}_{4}^{CCSDTQ-3}\contraction[0.5ex]{\left(\right.}{\hat{H}}{\left.e^{\hat{T}_{1}+\hat{T}_{2}}\left(1+\hat{T}_{3}\right)\right)|}{S}\left(\hat{H}e^{\hat{T}_{1}+\hat{T}_{2}}\left(1+\hat{T}_{3}\right)\right)_{c}|S\rangle\nonumber \\
 & +\langle0|\hat{\Lambda}_{3}\hat{V}\hat{T}_{4}^{CCSDTQ-3}|S\rangle\\
0 & =\langle0|(1+\hat{\Lambda})\contraction[0.5ex]{}{\bar{H}}{|}{D}\bar{H}|D\rangle_{CCSDT}+\langle0|\hat{\Lambda}_{4}^{CCSDTQ-3}\contraction[0.5ex]{\left(\right.}{\hat{H}}{\left.e^{\hat{T}_{1}+\hat{T}_{2}}\left(1+\hat{T}_{3}\right)\right)|}{D}\left(\hat{H}e^{\hat{T}_{1}+\hat{T}_{2}}\left(1+\hat{T}_{3}\right)\right)_{c}|D\rangle\\
0 & =\langle0|(1+\hat{\Lambda})\contraction[0.5ex]{}{\bar{H}}{|}{T}\bar{H}|T\rangle_{CCSDT}+\langle0|\hat{\Lambda}_{4}^{CCSDTQ-3}\contraction[1.0ex]{\left(\right.}{\hat{H}}{\left.e^{\hat{T}_{1}+\hat{T}_{2}}+\hat{P}_{T}\hat{V}\hat{T}_{3}\hat{P}_{D}\right)|}{T}\contraction[0.5ex]{\left(\hat{H}e^{\hat{T}_{1}+\hat{T}_{2}}+\hat{P}_{T}\!\!\right.}{\hat{V}}{\left.\hat{T}_{3}\hat{P}_{D}\right)|\!}{T}\left(\hat{H}e^{\hat{T}_{1}+\hat{T}_{2}}+\hat{P}_{T}\hat{V}\hat{T}_{3}\hat{P}_{D}\right)_{c}|T\rangle\nonumber \\
 & +\langle0|\contraction[0.5ex]{}{\hat{V}}{\left(\hat{P}_{D}\hat{\Lambda}_{4}^{CCSDTQ-3}\hat{T}_{3}\hat{P}_{T}\right)|}{T}\hat{V}\left(\hat{P}_{D}\hat{\Lambda}_{4}^{CCSDTQ-3}\hat{T}_{3}\hat{P}_{T}\right)|T\rangle
\end{align}

CC4:
\begin{align}
\hat{\Lambda}_{4}^{CC4} & =\left(\hat{\Lambda}_{2}\hat{V}+\hat{\Lambda}_{3}\left(\hat{H}+[\hat{V},\hat{T}_{1}]\right)\right)\hat{R}_{4}\\
0 & =\langle0|(1+\hat{\Lambda})\contraction[0.5ex]{}{\bar{H}}{|}{S}\bar{H}|S\rangle_{CCSDT}+\langle0|\hat{\Lambda}_{4}^{CC4}\contraction[0.5ex]{\left(\right.}{\hat{V}}{\left.\left(\hat{T}_{3}+\frac{1}{2}\hat{T}_{2}^{2}\right)\right)}{S}\left(\hat{V}^{\prime}\left(\hat{T}_{3}+\tfrac{1}{2}\hat{T}_{2}^{2}\right)\right)_{c}|S\rangle\nonumber \\
 & +\langle0|\hat{\Lambda}_{3}\contraction[0.5ex]{}{\hat{V}}{\hat{T}_{4}^{CC4}|}{S}\hat{V}\hat{T}_{4}^{CC4}|S\rangle\\
0 & =\langle0|(1+\hat{\Lambda})\contraction[0.5ex]{}{\bar{H}}{|}{D}\bar{H}|D\rangle_{CCSDT}+\langle0|\hat{\Lambda}_{4}^{CC4}\contraction[0.5ex]{[}{\hat{V}}{^{\prime},\hat{T}_{2}]|}{D}[\hat{V}^{\prime},\hat{T}_{2}]|D\rangle\\
0 & =\langle0|(1+\hat{\Lambda})\contraction[0.5ex]{}{\bar{H}}{|}{T}\bar{H}|T\rangle_{CCSDT}+\langle0|\hat{\Lambda}_{4}^{CC4}\hat{V}^{\prime}|T\rangle
\end{align}
where $\langle0|(1+\hat{\Lambda})\contraction[0.5ex]{}{\bar{H}}{|}{X}\bar{H}|X\rangle_{CCSDT}=\langle0|(1+\hat{\Lambda}_{1}+\hat{\Lambda}_{2}+\hat{\Lambda}_{3})\contraction[0.5ex]{}{\bar{H}}{_{CCSDT}|}{X}\bar{H}_{CCSDT}|X\rangle$
are the CCSDT $\hat{\Lambda}$ equations. As mentioned by Gauss et
al.,\cite{gaussAnalyticFirstSecond2000} the equations for the $\hat{\Lambda}$
amplitudes may also be derived diagrammatically by ``capping'' the
each cluster amplitude diagram with the appropriate $\hat{\Lambda}$
amplitude vertex to form a closed diagram, and then sequentially deleting
each cluster amplitude vertex to give the set of open $\hat{\Lambda}$
amplitude diagrams.

As with CCSDT(Q), the gradient and molecular properties are computed
through the one- and two-particle density matrices. These are easily
derived from (\ref{eq:approx-grad}),
\begin{align}
\left(D_{CCSDTQ-1a}\right)_{q}^{p} & =\left(D_{CCSDT}\right)_{q}^{p}-\delta_{pq}\langle0|\hat{\Lambda}_{4}^{CCSDTQ-1a}\{p^{\dagger}q\}\;\hat{T}_{4}^{CCSDTQ-1a}|0\rangle\\
\left(D_{CCSDTQ-1b}\right)_{q}^{p} & =\left(D_{CCSDT}\right)_{q}^{p}+\Delta_{pq}\langle0|\left(\hat{\Lambda}_{3}-\hat{\Lambda}_{4}^{CCSDTQ-1b}\right)\{p^{\dagger}q\}\;\hat{T}_{4}^{CCSDTQ-1b}|0\rangle\\
\left(D_{CCSDTQ-3}\right)_{q}^{p} & =\left(D_{CCSDT}\right)_{q}^{p}+\Delta_{pq}\langle0|\left(\hat{\Lambda}_{3}-\hat{\Lambda}_{4}^{CCSDTQ-3}\right)\{p^{\dagger}q\}\;\hat{T}_{4}^{CCSDTQ-3}|0\rangle\nonumber \\
 & +\langle0|\hat{\Lambda}_{4}^{CCSDTQ-3}\left(\{p^{\dagger}q\}\;\hat{T}_{3}\hat{T}_{2}\right)_{c}|0\rangle\\
\left(D_{CC4}\right)_{q}^{p} & =\left(D_{CCSDT}\right)_{q}^{p}+\Delta_{pq}\langle0|\left(\hat{\Lambda}_{3}-\hat{\Lambda}_{4}^{CC4}\right)\{p^{\dagger}q\}\;\hat{T}_{4}^{CC4}|0\rangle\\
\left(\Gamma_{CCSDTQ-1a}\right)_{rs}^{pq} & =\left(\Gamma_{CCSDT}\right)_{rs}^{pq}+\langle0|\hat{\Lambda}_{2}\;\{p^{\dagger}q^{\dagger}sr\}\;\hat{T}_{4}^{CCSDTQ-1a}|0\rangle\nonumber \\
 & +\langle0|\hat{\Lambda}_{4}^{CCSDTQ-1a}\left(\{p^{\dagger}q^{\dagger}sr\}\left(\hat{T}_{3}+\tfrac{1}{2}\hat{T}_{2}^{2}\right)\right)_{c}|0\rangle\\
\left(\Gamma_{CCSDTQ-1b}\right)_{rs}^{pq} & =\left(\Gamma_{CCSDT}\right)_{rs}^{pq}+\langle0|\hat{\Lambda}_{2}\;\{p^{\dagger}q^{\dagger}sr\}\;\hat{T}_{4}^{CCSDTQ-1b}|0\rangle\nonumber \\
 & +\langle0|\hat{\Lambda}_{3}\left(\{p^{\dagger}q^{\dagger}sr\}\;\left(1+\hat{T}_{1}\right)\hat{T}_{4}^{CCSDTQ-1b}\right)_{c}|0\rangle\nonumber \\
 & +\langle0|\hat{\Lambda}_{4}^{CCSDTQ-1b}\left(\{p^{\dagger}q^{\dagger}sr\}\left(\hat{T}_{3}+\tfrac{1}{2}\hat{T}_{2}^{2}\right)\right)_{c}|0\rangle\\
\left(\Gamma_{CCSDTQ-3}\right)_{rs}^{pq} & =\left(\Gamma_{CCSDT}\right)_{rs}^{pq}+\langle0|\hat{\Lambda}_{2}\;\{p^{\dagger}q^{\dagger}sr\}\,\hat{T}_{4}^{CCSDTQ-3}|0\rangle\nonumber \\
 & +\langle0|\hat{\Lambda}_{3}\left(\{p^{\dagger}q^{\dagger}sr\}\;\left(1+\hat{T}_{1}\right)\hat{T}_{4}^{CCSDTQ-1b}\right)_{c}|0\rangle\nonumber \\
 & +\langle0|\hat{\Lambda}_{4}^{CCSDTQ-1b}\left(\{p^{\dagger}q^{\dagger}sr\}\;e^{\hat{T}_{1}+\hat{T}_{2}}\left(1+\hat{T}_{3}\right)\right)_{c}|0\rangle\nonumber \\
 & +\langle0|\hat{\Lambda}_{4}^{CCSDTQ-1b}\left(\left(\hat{P}_{D}\;\{p^{\dagger}q^{\dagger}sr\}\;\hat{T}_{3}\hat{P}_{S}\right)\hat{T}_{3}\right)_{c}|0\rangle\\
\left(\Gamma_{CC4}\right)_{rs}^{pq} & =\left(\Gamma_{CCSDT}\right)_{rs}^{pq}+\langle0|\hat{\Lambda}_{2}\;\{p^{\dagger}q^{\dagger}sr\}\;\hat{T}_{4}^{CC4}|0\rangle\nonumber \\
 & +\langle0|\hat{\Lambda}_{3}\left(\{p^{\dagger}q^{\dagger}sr\}\,\left(1+\hat{T}_{1}\right)\hat{T}_{4}^{CC4}\right)_{c}|0\rangle\nonumber \\
 & +\langle0|\hat{\Lambda}_{4}^{CC4}\left(\{p^{\dagger}q^{\dagger}sr\}\;e^{\hat{T}_{1}}\left(\hat{T}_{3}+\tfrac{1}{2}\hat{T}_{2}^{2}\right)\right)_{c}|0\rangle
\end{align}
where $\left(D_{CCSDT}\right)_{q}^{p}=\langle0|(1+\hat{\Lambda}_{1}+\hat{\Lambda}_{2}+\hat{\Lambda}_{3})\left(\{p^{\dagger}q\}\;e^{\hat{T}_{1}+\hat{T}_{2}+\hat{T}_{3}}\right)|0\rangle$
and $\left(\Gamma_{CCSDT}\right)_{rs}^{pq}=\langle0|(1+\hat{\Lambda}_{1}+\hat{\Lambda}_{2}+\hat{\Lambda}_{3})\left(\{p^{\dagger}q^{\dagger}sr\}\;e^{\hat{T}_{1}+\hat{T}_{2}+\hat{T}_{3}}\right)|0\rangle$
are the CCSDT density matrices, but with cluster and $\hat{\Lambda}$
amplitudes determined by the modified equations for the specific method
in question. As with the $\hat{\Lambda}$ amplitudes, the density
matrices may be derived diagrammatically starting from the diagrammatic
form of the energy functional (\ref{eq:approx-ener}), and then deleting
the Hamiltonian vertex. Using the diagrammatic approach, care must
be taken with the sign and numerical prefactors such that the definition
of the density matrices matches that of (\ref{eq:cc-gradient-density}).

\section{Implementation}

The total CCSDT(Q), CCSDT(Q)/A, CCSDT(Q)/B, CCSDTQ-1a, CCSDTQ-1b,
CCSDTQ-3, and CC4 density matrices have been implemented in the NCC
module\cite{matthewsNonorthogonalSpinadaptationCoupled2015} of the
CFOUR program system.\cite{stantonCFOURCoupledClusterTechniques}
As of this time, this implementation only handles closed-shell reference
states, including RHF and closed-shell QRHF (or other restricted non-HF
orbitals, e.g. Kohn-Sham orbitals). Additionally, the implementation
makes use of non-orthogonal spin-adaptation techniques\cite{matthewsNonorthogonalSpinadaptationCoupled2015,matthewsChapter10Diagrams2019}
as well as recent advances in high-performance tensor contraction,\cite{matthewsHighPerformanceTensorContraction2018,matthewsExtendingOptimisingDirect2019}
which allows for high efficiency and relatively compact working equations.

The cost of a CCSDT(Q) gradient calculation is between two and three
times that of the corresponding energy calculation. The energy calculation
requires the iterative solution of the CCSDT coupled cluster equations
with $\mathscr{O}(n^{8})$ cost, and four non-iterative $\mathscr{O}(n^{9})$
steps (multiplication of $\hat{T}_{3}$ with $v_{ci}^{ab}$ and $v_{jk}^{ia}$,
and $\hat{T}_{2}$ with the three-particle intermediates $\tilde{W}_{ije}^{abc}$
and $\tilde{W}_{ijk}^{abm}$ {[}see below{]}). The gradient calculation
instead requires two iterative $\mathscr{O}(n^{8})$ procedures (the
CCSDT coupled cluster equations and the combined $\hat{\Lambda}+\tilde{\Lambda}$
equations) and ten non-iterative $\mathscr{O}(n^{9})$ steps. In order
to reach this minimal number of $\mathscr{O}(n^{9})$ steps, the contributions
to $(\Gamma_{(Q)}^{\prime})_{ci}^{ab}$ and $(\Gamma_{(Q)}^{\prime})_{ab}^{ci}$
must be combined (and similarly for $(\Gamma_{(Q)}^{\prime})_{jk}^{ia}$
and $(\Gamma_{(Q)}^{\prime})_{ia}^{jk}$) when computing the symmetrized
two-electron density,
\begin{align}
\left(\Gamma_{(Q)}^{\prime}\right)_{ci}^{ab} & =\left(\Gamma_{(Q)}^{\prime}\right)_{ab}^{ci}=\frac{1}{24}\sum_{efmno}\left\{ (t^{[3]})_{mnoi}^{efab}+(t^{\prime})_{mnoi}^{efab}\right\} t_{mno}^{efc}\\
\left(\Gamma_{(Q)}^{\prime}\right)_{jk}^{ia} & =\left(\Gamma_{(Q)}^{\prime}\right)_{ia}^{jk}=-\frac{1}{24}\sum_{efgmn}\left\{ (t^{[3]})_{mnjk}^{efga}+(t^{\prime})_{mnjk}^{efga}\right\} t_{mni}^{efg}
\end{align}
In the same way, the $\hat{T}_{4}^{[3]}$ and $T_{4}^{\prime}$ contributions
to $\hat{S}_{3}$ may be computed at the same time. All other energy
contributions and density matrix elements scale as $\mathscr{O}(n^{8})$
or less.

For the CCSDT(Q)/A and CCSDT(Q)/B energies and gradients, the cost
is increased somewhat over CCSDT(Q). In the case of CCSDT(Q)/B, the
introduction of $\hat{T}_{3}^{B}$ increases the number of $\mathscr{O}(n^{9})$
steps to six for the energy, and to 16 for the gradient, since the
density matrix contributions for $\hat{T}_{3}$ and $\hat{T}_{3}^{B}$
must be computed separately, and similarly for the contributions to
$\hat{S}_{3}^{B}$. For CCSDT(Q)/A, the cost is the same except that
the $\hat{T}_{1}^{\dagger}\hat{T}_{2}^{\dagger}$ term could in theory
be factorized such that it is computed at only $\mathscr{O}(n^{8})$
cost, but to the authors' knowledge this has not been done in practice.
The cost of a CCSDT(Q)/A,B energy or gradient is therefore intermediate
between CCSDT(Q) and CCSDT(Q)$_{\Lambda}$, where the latter requires
the same number of $\mathscr{O}(n^{9})$ steps as CCSDT(Q)/B but twice
as many iterative $\mathscr{O}(n^{8})$ equations. The analytic gradients
of CCSDT(Q)$_{\Lambda}$ will be studied in detail in a later publication.

In both CCSDT(Q) and CCSDT(Q)/A,B, the $\hat{T}_{4}^{[3],\prime,A,B}$
amplitudes need not be stored and may be computed on the fly for evaluation
of the energy and contraction into the density matrices and $\hat{S}$
intermediates. However, unlike in CCSD(T), where no additional storage
beyond that required for CCSD is needed, CCSDT(Q) does require the
calculation of three-particle intermediates,
\begin{align}
\tilde{W}_{ije}^{abc} & =\frac{1}{2}P(a/bc)\sum_{f}v_{fe}^{bc}t_{ij}^{af}\\
\tilde{W}_{ijk}^{abm} & =-\frac{1}{2}P(i/jk)\sum_{n}v_{jk}^{nm}t_{in}^{ab}+P(a/b)P(ij/k)\sum_{e}v_{ek}^{bm}t_{ij}^{ae}
\end{align}
where the permutation operator $P$ antisymmetrizes the labels on
either side of the slash. For CCSDT(Q) gradients, the similar intermediate
three-particle density matrices $\tilde{\Gamma}_{bcd}^{aij}$ and
$\tilde{\Gamma}_{abl}^{ijk}$ are also required. These intermediates
must either be stored on disk or recalculated as needed. In our implementation
we have found that storing them to disk tends to be the most efficient
solution as recalculation is often more expensive and can, in extreme
circumstances, even increase the formal scaling of the this term.
However, as the $\tilde{W}_{ije}^{abc}$ intermediate in particular
may be much larger than even the $\hat{T}_{3}$ amplitudes, the disk
space requirements of CCSDT(Q) calculations are indeed more strenuous
than for CCSDT.

The iterative approximations all scale as $\mathscr{O}(n^{9})$, with
the rough order of cost given by $\text{CCSDTQ-3}>\text{CC4}>\text{CCSDTQ-1b}>\text{CCSDTQ-1a}$.
The main feature that impacts the computational cost is the number
of $\mathscr{O}(n^{9})$ steps. For CCSDTQ-1a, the cluster amplitude
equations require only four such steps per iteration, while all other
methods require six steps due to the direct $\hat{T}_{4}\rightarrow\hat{T}_{3}$
coupling. The difference in the $\hat{\Lambda}$ equations is even
larger: CCSDTQ-1a again requires four $\mathscr{O}(n^{9})$ steps,
but the other methods now require 12 steps per iteration due to the
fact the the $\hat{T}_{4}$ amplitudes are required in addition to
$\hat{\Lambda}_{4}$, and that these amplitudes are generally recomputed
rather than stored due to their extremely large size. Additionally,
the current implementations of the CCSDTQ-3 and CC4 $\hat{\Lambda}$
equations are sub-optimal in that they include two additional $\mathscr{O}(n^{9})$
steps per iteration. The construction of the density matrices is not
a major bottleneck for the iterative approximations because it occurs
only once and not every iteration.

Many terms in the CCSDTQ-3 amplitude equations may be included by
defining suitable two- and three-particle intermediates (including
the seemingly expensive $(\hat{P}_{D}\hat{V}\hat{T}_{3}\hat{P}_{S})\hat{T}_{3}$
term), but in particular the terms $t_{ijkl}^{abcd}\gets v_{ef}^{mn}t_{ij}^{ae}t_{k}^{f}t_{mnl}^{bcd}-v_{ef}^{mn}t_{ij}^{ae}t_{m}^{b}t_{nkl}^{fcd}$,
while only scaling as $\mathscr{O}(n^{8})$ involve contributions
from $\hat{T}_{3}$ into an intermediate $\tilde{W}_{ije}^{abc}$.
The large size of the inputs and outputs and relatively small size
of the summation indices leads to low efficiency and a noticeably
increased cost for CCSDTQ-3 compared to CCSDTQ-1b and CC4. The same
is true of the contributions from a three-particle intermediate $\tilde{\Gamma}_{bcd}^{aij}$
to the CCSDTQ-3 $\hat{\Lambda}_{3}$ equations. The iterative methods
may also benefit from convergence acceleration via sub-iteration,\cite{matthewsAcceleratingConvergenceHigherorder2015,doi:10.1080/00268976.2020.1757774}
i.e. holding the $\hat{T}_{4}$ and/or $\hat{\Lambda}_{4}$ amplitudes
constant while iteratively improving the other amplitudes. The contributions
from the quadruples amplitudes may then be computed once per outer
iteration and then added in at low cost during the inner iterations.
In this work, sub-iterations is included in the cluster amplitude
equations but not for the $\hat{\Lambda}$ amplitudes.

The accuracy of the present equations and implementation has been
checked by comparing the computed gradients to finite differences
of energies up to approximately $10^{-9}$ relative error in the molecular
gradient where the finite difference method reaches the limit of its
accuracy. The single-point energies have also been checked against
the MRCC program of Kállay.\cite{kallayMRCCProgramSystem2020} The
construction of the one- and two-particle density matrices has also
been checked by contracting them with the Hamiltonian to reproduce
the various energies and/or energy corrections to numerical accuracy.
A final check for the non-iterative approximations is contracting
the $\hat{S}$ amplitudes with the corresponding $\hat{T}$ amplitudes,
which should give the sum of all energy terms that include that $\hat{T}$
amplitude (perhaps with some redundancy). In CCSDT(Q) for example,
$\langle0|\hat{S}_{2}\hat{T}_{2}|0\rangle=\langle0|\hat{T}_{2}^{\dagger}\hat{V}\hat{T}_{4}^{[3]}|0\rangle+\langle0|\hat{T}_{4}^{\prime}\left(\hat{V}\hat{T}_{2}^{2}\right)_{c}|0\rangle$.
The correctness of these expressions has also been checked.

\section{Results and Discussion}

The new implementations of approximate quadruples methods in CFOUR
have been applied to two prototypical test systems: the isomerization
of dimethylcarbene (DMC) to propene, and the simplest Criegee Intermediate
(CI),\cite{criegeeOzonisierung10Oktalins1949} $\ce{H2COO}$. In both
cases, the optimized equilibrium and transition state structures and
harmonic vibrational frequencies (via finite differences of gradients)
have been determined at each level of theory, as well as with CCSD(T)
and CCSDT for comparison. Absolute and relative (for the DMC--propene
system) energies including harmonic vibrational zero-point energy
were also calculated. In order to obtain an internally consistent
benchmark and avoid complications with respect to basis set convergence,
core correlation, relativistic effects, spin-orbit coupling, anharmonicity,
etc., the computed values are compared against full CCSDTQ calculations
instead of to experimental values. Core electrons are frozen in all
calculations, and the double-$\zeta$ truncation of the ANO basis
set of Almlöf and Taylor,\cite{almlofAtomicNaturalOrbital1991} often
dubbed ANO0, is used. SCF, CC/$\Lambda$, and geometry optimization
thresholds were set to $10^{-10}$, $10^{-9}$, and $10^{-8}$, respectively,
with all other settings set to default. Harmonic vibrational frequencies
for translational and rotational modes are below 0.2 cm$^{-1}$ in
all cases. 

\subsection{\label{subsec:Dimethylcarbene-Isomerization}Dimethylcarbene Isomerization}

\begin{table}
\begin{centering}
\subfloat[$g$-DMC]{\begin{centering}
\begin{tabular}{|c||r@{\extracolsep{0pt}.}l|r@{\extracolsep{0pt}.}l|r@{\extracolsep{0pt}.}l|r@{\extracolsep{0pt}.}l|r@{\extracolsep{0pt}.}l|r@{\extracolsep{0pt}.}l|r@{\extracolsep{0pt}.}l|r@{\extracolsep{0pt}.}l|r@{\extracolsep{0pt}.}l|}
\hline 
 & \multicolumn{2}{c|}{(T)} & \multicolumn{2}{c|}{T} & \multicolumn{2}{c|}{(Q)} & \multicolumn{2}{c|}{(Q)/A} & \multicolumn{2}{c|}{(Q)/B} & \multicolumn{2}{c|}{Q-1a} & \multicolumn{2}{c|}{Q-1b} & \multicolumn{2}{c|}{Q-3} & \multicolumn{2}{c|}{CC4}\tabularnewline
\hline 
\hline 
$\Delta ABC$ & 0&09\% & 0&03\% & 0&00\% & 0&00\% & 0&00\% & 0&01\% & 0&00\% & 0&01\% & 0&00\%\tabularnewline
\hline 
$\Delta r$ & 0&0003 & 0&0001 & 0&0000 & 0&0000 & 0&0000 & 0&0001 & 0&0000 & 0&0001 & 0&0000\tabularnewline
\hline 
$\Delta r_{\text{HH}}$ & -0&0034 & -0&0003 & -0&0002 & -0&0001 & -0&0001 & 0&0000 & -0&0002 & -0&0001 & -0&0002\tabularnewline
\hline 
$\Delta\angle$,$\Delta\phi$ & 0&087 & 0&003 & 0&006 & 0&002 & 0&004 & 0&014 & 0&006 & 0&002 & 0&006\tabularnewline
\hline 
$\Delta\omega$ & 2&11 & 1&06 & 0&11 & 0&05 & 0&04 & 0&97 & 0&05 & 0&47 & 0&07\tabularnewline
\hline 
$\Delta E_{0}$ & 1&337 & 0&507 & -0&017 & 0&019 & 0&006 & 0&394 & 0&009 & 0&198 & 0&005\tabularnewline
\hline 
$\Delta HVZPE$ & \hphantom{-}0&059 & \hphantom{-}0&032 & -0&003 & \hphantom{-}0&001 & \hphantom{-}0&000 & \hphantom{-}0&029 & \hphantom{-}0&000 & \hphantom{-}0&014 & -0&001\tabularnewline
\hline 
\end{tabular}
\par\end{centering}
}
\par\end{centering}
\begin{centering}
\subfloat[$g$-TS]{\begin{centering}
\begin{tabular}{|c||r@{\extracolsep{0pt}.}l|r@{\extracolsep{0pt}.}l|r@{\extracolsep{0pt}.}l|r@{\extracolsep{0pt}.}l|r@{\extracolsep{0pt}.}l|r@{\extracolsep{0pt}.}l|r@{\extracolsep{0pt}.}l|r@{\extracolsep{0pt}.}l|r@{\extracolsep{0pt}.}l|}
\hline 
 & \multicolumn{2}{c|}{(T)} & \multicolumn{2}{c|}{T} & \multicolumn{2}{c|}{(Q)} & \multicolumn{2}{c|}{(Q)/A} & \multicolumn{2}{c|}{(Q)/B} & \multicolumn{2}{c|}{Q-1a} & \multicolumn{2}{c|}{Q-1b} & \multicolumn{2}{c|}{Q-3} & \multicolumn{2}{c|}{CC4}\tabularnewline
\hline 
\hline 
$\Delta ABC$ & 0&07\% & 0&04\% & 0&01\% & 0&00\% & 0&00\% & 0&03\% & 0&00\% & 0&02\% & 0&00\%\tabularnewline
\hline 
$\Delta r$ & 0&0004 & 0&0003 & 0&0001 & 0&0000 & 0&0000 & 0&0002 & 0&0000 & 0&0001 & 0&0000\tabularnewline
\hline 
$\Delta\angle$,$\Delta\phi$ & 0&018 & 0&020 & 0&010 & 0&004 & 0&006 & 0&010 & 0&004 & 0&007 & 0&005\tabularnewline
\hline 
$\Delta\omega$ & 2&41 & 1&24 & 0&22 & 0&12 & 0&13 & 1&12 & 0&11 & 0&57 & 0&13\tabularnewline
\hline 
$\Delta E_{0}$ & 1&201 & 0&562 & -0&045 & -0&002 & -0&017 & 0&435 & -0&011 & 0&221 & -0&016\tabularnewline
\hline 
$\Delta T_{0}$ & -0&136 & 0&055 & -0&028 & -0&021 & -0&023 & 0&041 & -0&020 & 0&023 & -0&022\tabularnewline
\hline 
$\Delta HVZPE$ & \hphantom{-}0&061 & \hphantom{-}0&036 & -0&006 & \hphantom{-}0&001 & -0&002 & \hphantom{-}0&032 & -0&001 & \hphantom{-}0&016 & -0&002\tabularnewline
\hline 
\end{tabular}
\par\end{centering}
}
\par\end{centering}
\begin{centering}
\subfloat[Propene]{\begin{centering}
\begin{tabular}{|c||r@{\extracolsep{0pt}.}l|r@{\extracolsep{0pt}.}l|r@{\extracolsep{0pt}.}l|r@{\extracolsep{0pt}.}l|r@{\extracolsep{0pt}.}l|r@{\extracolsep{0pt}.}l|r@{\extracolsep{0pt}.}l|r@{\extracolsep{0pt}.}l|r@{\extracolsep{0pt}.}l|}
\hline 
 & \multicolumn{2}{c|}{(T)} & \multicolumn{2}{c|}{T} & \multicolumn{2}{c|}{(Q)} & \multicolumn{2}{c|}{(Q)/A} & \multicolumn{2}{c|}{(Q)/B} & \multicolumn{2}{c|}{Q-1a} & \multicolumn{2}{c|}{Q-1b} & \multicolumn{2}{c|}{Q-3} & \multicolumn{2}{c|}{CC4}\tabularnewline
\hline 
\hline 
$\Delta ABC$ & 0&06\% & 0&04\% & 0&01\% & 0&01\% & 0&01\% & 0&02\% & 0&00\% & 0&01\% & 0&01\%\tabularnewline
\hline 
$\Delta r$ & 0&0002 & 0&0002 & 0&0000 & 0&0000 & 0&0000 & 0&0001 & 0&0000 & 0&0001 & 0&0000\tabularnewline
\hline 
$\Delta\angle$,$\Delta\phi$ & 0&003 & 0&007 & 0&002 & 0&001 & 0&001 & 0&004 & 0&001 & 0&002 & 0&001\tabularnewline
\hline 
$\Delta\omega$ & 1&59 & 1&17 & 0&25 & 0&16 & 0&19 & 0&77 & 0&15 & 0&43 & 0&18\tabularnewline
\hline 
$\Delta E_{0}$ & 1&114 & 0&588 & -0&060 & -0&041 & -0&047 & 0&339 & -0&036 & 0&192 & -0&041\tabularnewline
\hline 
$\Delta T_{0}$ & -0&222 & 0&081 & -0&044 & -0&059 & -0&054 & -0&055 & -0&045 & -0&006 & -0&046\tabularnewline
\hline 
$\Delta HVZPE$ & 0&048 & 0&035 & -0&008 & -0&005 & -0&006 & 0&023 & -0&005 & 0&013 & -0&005\tabularnewline
\hline 
\end{tabular}
\par\end{centering}
}
\par\end{centering}
\caption{\label{tab:dmc-propene}Statistical errors in geometric, vibrational,
and energetic properties for \emph{gauche-}dimethylcarbene (\emph{g}-DMC),
propene, and the hydrogen-shift transition state (\emph{g}-TS). Errors
in equilibrium rotational constants ($\Delta ABC$) are listed as
mean average percent errors (MAPE), while other values are listed
as mean absolute errors (MAE). Bond length errors ($\Delta r$) are
in $\text{Å}$, bond and dihedral angle errors ($\Delta\angle,\Delta\phi$)
are in degrees, harmonic frequency errors ($\Delta\omega$) are in
cm$^{-1}$, and total ($\Delta E_{0}$), relative ($\Delta T_{0}$),
and harmonic vibrational zero-point energy $($$\Delta HVZPE$) errors
are in kcal/mol. Total and relative energies include zero-point energy.
The theoretical methods are listed in an abbreviated notation, e.g.
Q-1a = CCSDTQ-1a.}
\end{table}

The most stable conformation of singlet DMC is the C$_{2}$\emph{
gauche} conformation ($g$-DMC),\cite{richardsDimethylcarbeneSingletGround1995}
and the DMC and propene minima are connected by a single chiral transition
state ($g$-TS) along the 1,2--hydrogen shift isomerization pathway.
Table \ref{tab:dmc-propene} lists the mean absolute errors for various
categories of geometrical, energetic, and vibrational quantities compared
to CCSDTQ. At each geometry, the error in the rotational constants
rapidly decreases from 0.06--0.09\% down to essentially zero when
full triple and then quadruple excitations are included. The same
is true of the errors in bond lengths, although these are already
fairly small at the CCSD(T) level ($\SI{\sim30}{\femto\m}$). An exception
is the non-bonded \emph{gauche} H--H distance in DMC ($\Delta r_{\text{HH}}$),
which is quite sensitive to higher-order correlation effects. The
bond and dihedral angle errors similarly tell a different story: the
angle error at the CCSD(T) level increases dramatically from propene
backwards along the isomerization pathway. For DMC, the mean absolute
error is nearly $\ang{0.1}$, and still $\ang{0.02}$ at the transition
state. For both DMC and $g$-TS, the angle error for the approximate
quadruples methods is also higher than for propene by a factor of
roughly five. Interestingly, CCSDT does not improve on CCSD(T) by
this measure for propene or the transition state, but in contrast
performs fully as well as the approximate quadruples methods for DMC.
The high angle error in the DMC structures, along with the spuriously
large non-bonded H--H distance seem to indicate that the major geometrical
errors in DMC relate to non-bonding intermolecular interactions. As
these interactions arise primarily form many-body effects, the fact
that connected triple excitations (at least) are necessary to fully
describe the structure is not surprising. An alternative hypothesis
is that the biradical nature of DMC could account for the greater
importance of higher-order correlation. However, a better description
of biradical character would be expected to primarily affect the C--C--C
angle. An inspection of the detailed data in the Supplemental Information
shows that this is not the case, and that instead the largest angle
errors are involved in the reorganization of the methyl hydrogens,
especially the \emph{gauche} hydrogen. This suggests that very high-accuracy
structures with steric crowding (and potentially other strains such
as ring strain) would benefit from the inclusion of quadruple excitations.
Non-bonded intermolecular interactions are another area of possible
applicability. Lane et al. calculated the quadruples contribution
to the equilibrium geometry of the water dimer,\cite{laneCCSDTQOptimizedGeometry2013}
and while the effect on the intermolecular distance was small, there
was a somewhat larger effect on the relative angles of the two water
monomers (predominately the acceptor wag angle).

The harmonic vibrational frequencies for all three structures show
a roughly order-of-magnitude reduction in error for the approximate
quadruples methods (with the exception of CCSDTQ-1a), while CCSDT
roughly halves the error with respect to CCSD(T). Since CCSD(T) is
in error approximately $\SI{2}{\per\cm}$, the inclusion of quadruple
excitations is critical to achieving spectroscopic (sub--cm$^{-1}$)
accuracy. While inspection of the detailed results in the Supplemental
Information shows that there are not any particular vibrational modes
with excessively large error for DMC or propene, two modes of the
transition state show errors in excess of $\SI{5}{\per\cm}$ at the
CCSD(T) level, not surprisingly both involving movement of the migratory
hydrogen. A highly-accurate description of these vibrational modes
is essential to the description of kinetic tunneling, e.g. through
semi-classical transition state theory.\cite{millerSemiclassicalLimitQuantum1975,millerQuantumSemiclassicalTheory1998,nguyenPracticalImplementationSemiclassical2010}
Errors in the absolute energies are rather uninteresting, falling
from just above 1 kcal/mol down to approximately 0.2 kJ/mol for the
most accurate approximate quadruples methods (keep in mind that this
is with respect to CCSDTQ, and errors compared to experiment will
be larger). Outliers are CCSDTQ-1a and, more surprisingly, CCSDTQ-3,
which quadruple or more the error of other approximate quadruples
methods on average. Also to note is that the A and B variants of CCSDT(Q),
while originally designed for open-shell ROHF calculations, also seem
to slightly improve on ``normal'' CCSDT(Q) for these closed-shell
examples. When examining relative errors ($\Delta T_{0}$) for the
transition state and propene, CCSD(T) is seen to benefit greatly from
error cancellation, bringing the average error down to 0.6--0.9 kJ/mol.
While CCSDT similarly benefits from error cancellation, albeit to
a lesser degree, only CCSDTQ-1a and CCSDTQ-3 gain any error cancellation
benefit amongst the approximate quadruples methods. While this cancellation
brings them in line with the other methods when considering relative
energies, the comparatively poor behavior for absolute energies is
rather troublesome and CCSDTQ-1a in particular should likely not be
considered reliable.

\subsection{\label{subsec:Criegee-Intermediate}Criegee Intermediate}

\begin{table}
\begin{centering}
\begin{tabular}{|c||r@{\extracolsep{0pt}.}l|r@{\extracolsep{0pt}.}l|r@{\extracolsep{0pt}.}l|r@{\extracolsep{0pt}.}l|r@{\extracolsep{0pt}.}l|r@{\extracolsep{0pt}.}l|r@{\extracolsep{0pt}.}l|r@{\extracolsep{0pt}.}l|r@{\extracolsep{0pt}.}l|}
\hline 
 & \multicolumn{2}{c|}{(T)} & \multicolumn{2}{c|}{T} & \multicolumn{2}{c|}{(Q)} & \multicolumn{2}{c|}{(Q)/A} & \multicolumn{2}{c|}{(Q)/B} & \multicolumn{2}{c|}{Q-1a} & \multicolumn{2}{c|}{Q-1b} & \multicolumn{2}{c|}{Q-3} & \multicolumn{2}{c|}{CC4}\tabularnewline
\hline 
\hline 
$\Delta ABC$ & 0&34\% & 0&27\% & 0&72\% & 0&26\% & 0&35\% & 0&33\% & 0&04\% & 0&11\% & 0&04\%\tabularnewline
\hline 
$\Delta r$ & 0&0013 & 0&0018 & 0&0025 & 0&0011 & 0&0014 & 0&0013 & 0&0002 & 0&0004 & 0&0004\tabularnewline
\hline 
$\Delta\angle$,$\Delta\phi$ & 0&063 & 0&074 & 0&161 & 0&127 & 0&082 & 0&094 & 0&023 & 0&017 & 0&014\tabularnewline
\hline 
$\Delta\omega$ & 7&41 & 10&32 & 9&43 & 3&38 & 4&97 & 10&01 & 1&03 & 2&29 & 1&84\tabularnewline
\hline 
$\Delta E_{0}$ & 2&117 & 1&878 & -1&185 & 0&310 & -0&371 & 1&306 & -0&009 & 0&448 & -0&226\tabularnewline
\hline 
$\Delta HVZPE$ & 0&060 & 0&089 & -0&102 & 0&021 & -0&049 & 0&076 & 0&006 & 0&024 & -0&011\tabularnewline
\hline 
\end{tabular}
\par\end{centering}
\caption{\label{tab:criegee}Statistical errors in geometric, vibrational,
and energetic properties for CI. Errors in equilibrium rotational
constants ($\Delta ABC$) are listed as mean average percent errors
(MAPE), while other values are listed as mean absolute errors (MAE).
Bond length errors ($\Delta r$) are in $\text{Å}$, bond and dihedral
angle errors ($\Delta\angle,\Delta\phi$) are in degrees, harmonic
frequency errors ($\Delta\omega$) are in cm$^{-1}$, and total ($\Delta E_{0}$)
and harmonic vibrational zero-point energy $($$\Delta HVZPE$) errors
are in kcal/mol. The total energy includes zero-point energy. The
theoretical methods are listed in an abbreviated notation, e.g. Q-1a
= CCSDTQ-1a.}
\end{table}

Mean absolute errors for CI are given in Table \ref{tab:criegee}.
Right off the bat, an inspection of the errors in the rotational constants,
with errors approximately five times larger, show that CI is significantly
more sensitive to higher-order correlation effects than the DMC--propene
system. Increases in error of a similar magnitude are evident for
bond lengths and angles, except that the angle error of DMC still
outstrips that of CI at the CCSD(T) level. A major difference between
DMC--propene and CI is that a number of higher-order methods fail
to improve on CCSD(T) for these geometric quantities, with only CCSDTQ-1b
and CC4 (and to some extent CCSDTQ-3) providing a reliable and significant
decrease in error. Interestingly, CCSDT(Q) dramatically worsens all
geometric errors by a factor of two; CCSDT(Q)/A and B improve somewhat
but do not reach the accuracy of CCSDTQ-1b or CC4. In particular,
the O--O bond distance is vastly improved on going from CCSD(T) to
the more accurate quadruples methods. On the other hand, the C--O
bond distance, which is well reproduced by CCSD(T), is problematic
for CCSDT(Q) (but not A and B) and CCSDTQ-1a, and even for CCSDT.
Among the angles, the O--O--C angle is clearly the most problematic,
although several approximate quadruples methods as well as CCSDT worsen
the O--C--H bond angles compared to CCSD(T). These results suggest
that non-iterative approximations of quadruple excitations may not
be able to reliably improve on CCSD(T) for describing the geometric
parameters of moderately multi-configurational systems.

The harmonic frequencies show a similarly disappointing pattern, with
CCSDT, CCSDT(Q), and CCSDTQ-1a showing a deterioration compared to
the CCSD(T) values. Here, CCSDT(Q)/A and B are able to improve upon
CCSD(T), with the A variant slightly in the lead. CCSDTQ-1b and CC4,
as before, again show a reliable reduction in error, although not
to the sub--cm$^{-1}$ level achieved for DMC--propene. A closer
inspection of the data in the Supplemental Information shows that
two frequencies, $\omega_{4}$ and $\omega_{6}$ are responsible for
the lion's share of the error, especially at the CCSD(T) level. These
frequencies correspond to C--O and O--O stretching vibrations. CCSDT
and beyond tend to show an improvement in O--O stretching, much as
they show an improvement in the O--O bond length, while C--O stretching
is worsened at the CCSDT level and to a slightly lesser extent at
the CCSDT(Q) level. CCSDT, CCSDT(Q), and CCSDTQ-1a further exhibit
severe errors (up to 30 cm$^{-1}$) in the out-of-plane motions that
are not present at the CCSD(T) level. It is quite difficult to ascribe
these errors to a particular feature of the theory, for example using
diagrammatic or perturbation arguments. Perhaps the simplest, albeit
the least satisfying, explanation is a tendency of some of the methods,
in particular CCSDT(Q), to ``overshoot'' the full CCSDTQ results.
This can be seen explicitly in the absolute energies, where CCSDT(Q)
overshoots CCSDTQ by more than a kcal/mol. On the other hand, CCSDT(Q)/A
and /B under- and overestimate the CCSDTQ contribution by roughly
the same amount, while CC4 also overshoots CCSDTQ. Thus, the extrapolation
of simple energetics to geometric and vibrational parameters is qualitative
at best. Perhaps the best example of the disconnect between energetics
and geometry is seen for CCSDT(Q), which, as noted above, performs
rather poorly for geometries and frequencies, but also halves the
error in the absolute energy compared to CCSD(T). CCSDTQ-3 does make
some improvement over CCSD(T) for CI, but is again not as accurate
as CCSDTQ-1b.

\subsection{Timings}

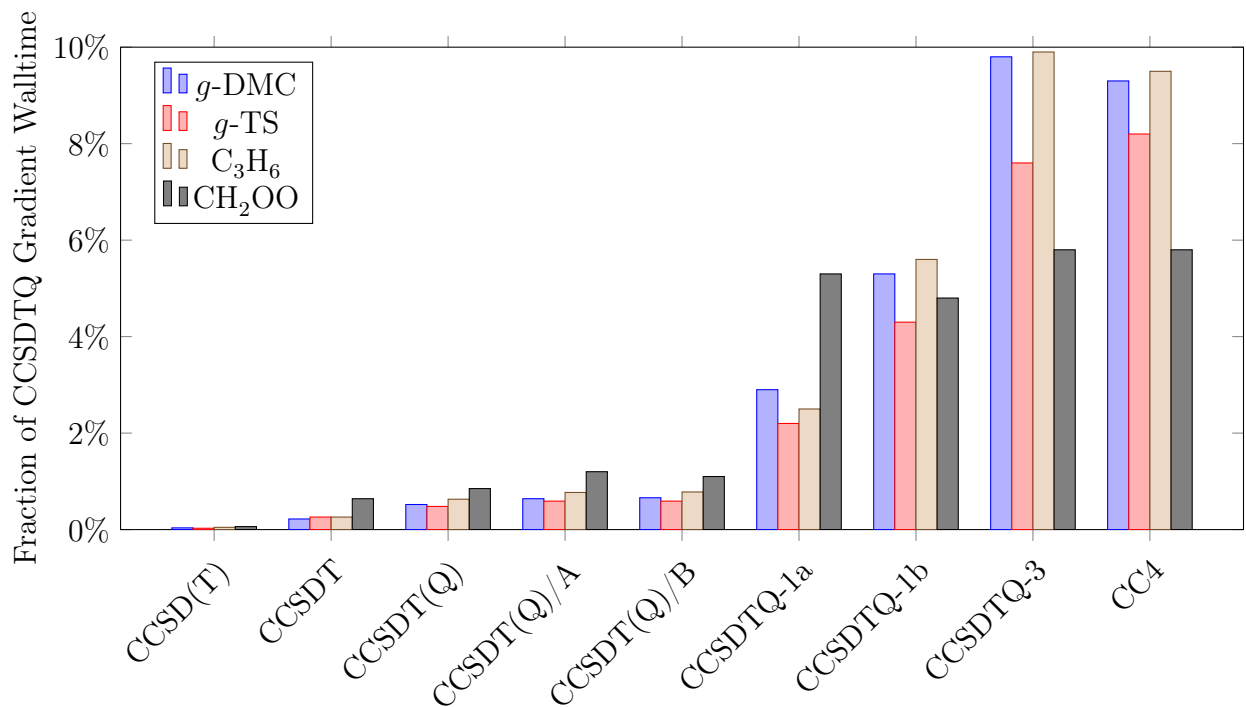
\begin{figure}
\begin{tikzpicture}
\begin{axis}[
ybar=0,
bar width=8,
ymin=0,
ymax=0.1,
yticklabel={\pgfmathparse{\tick*100}\pgfmathprintnumber{\pgfmathresult}\%},
width=\textwidth,
height=8cm,
xtick=data,
enlarge x limits=.1,
xticklabels={CCSD(T), CCSDT, CCSDT(Q), CCSDT(Q)/A, CCSDT(Q)/B, CCSDTQ-1a, CCSDTQ-1b, CCSDTQ-3, CC4},
legend columns=1,
ylabel={Fraction of CCSDTQ Gradient Walltime},
legend style={legend pos=north west},
x tick label style={rotate=45,anchor=north east}
]

\addplot coordinates {
(1,3.6E-04)	(2,2.2E-03)	(3,5.2E-03)	(4,6.4E-03)	(5,6.6E-03)	(6,2.9E-02)	(7,5.3E-02)	(8,9.8E-02)	(9,9.3E-02) 
};
\addplot coordinates {
(1,2.8E-04)	(2,2.6E-03)	(3,4.8E-03)	(4,5.9E-03)	(5,5.9E-03)	(6,2.2E-02)	(7,4.3E-02)	(8,7.6E-02)	(9,8.2E-02) 
};
\addplot coordinates {
(1,4.7E-04)	(2,2.6E-03)	(3,6.3E-03)	(4,7.7E-03)	(5,7.8E-03)	(6,2.5E-02)	(7,5.6E-02)	(8,9.9E-02)	(9,9.5E-02) 
};
\addplot coordinates {
(1,6.2E-04)	(2,6.4E-03)	(3,8.5E-03)	(4,1.2E-02)	(5,1.1E-02)	(6,5.3E-02)	(7,4.8E-02)	(8,5.8E-02)	(9,5.8E-02)
};
\legend{$g$-DMC, $g$-TS, \ce{C3H6}, \ce{CH2OO}}

\end{axis}
\end{tikzpicture}

\caption{\label{fig:timings}Timings (wall-time), relative to full CCSDTQ,
for one gradient evaluation for all calculations presented in \subsecref{Dimethylcarbene-Isomerization}
and \subsecref{Criegee-Intermediate}. All calculations utilized $2\times$Intel®
Xeon E5-2695v4 2.1 GHz processors (36 cores total) with $\SI{256}{\gibi\byte}$
RAM.}

\end{figure}

While the results above show that certain iterative, and in some cases,
non-iterative approximations to CCSDTQ can yield substantial accuracy
gains compared to CCSD(T) and CCSDT, the computational cost of these
methods is a critically important factor in determining when such
methods may be used. Full CCSDTQ scales with the tenth power of system
size ($\mathscr{O}(n^{10})$), and is applicable only to very small
molecular systems, especially when analytic gradients or properties
are desired. While the approximate methods described here reduce this
scaling to $\mathscr{O}(n^{9})$, they still represent a quite considerable
cost increase over CCSDT and especially CCSD(T). Figure \ref{fig:timings}
shows timings for a single gradient evaluation, as a percentage of
the CCSDTQ timing for all methods considered. CCSD(T) only requires
$\sim0.05\%$ of the time compared to CCSDTQ, while CCSDT requires
$\sim0.4\%$ of the CCSDTQ time. However, as CCSDT often fails to
improve over CCSD(T), it is necessary to include approximate quadruple
excitations if higher accuracy is required. CCSDT(Q) only increases
the computational time over CCSDT modestly, keeping the relative timing
under 1\%. The A and B variants, while increasing the cost again slightly,
likewise remain in the 1\% range. When these methods can be considered
reliable (strongly single-configurational systems with important many-body
dynamic correlation), this modest increase in time over CCSDT for
systems of the size studied here shows that they can be applied rather
routinely. For CI, however, the non-iterative approximations are not
reliable and an iterative approximations is mandatory. Of the iterative
methods, CCSDTQ-1a is the cheapest, requiring 3--4\% of the CCSDTQ
time. As shown above, however, this methods consistently fails to
improve upon CCSD(T). CCSDTQ-1b, however, ``only'' increases the
computational time to 4--5\% of CCSDTQ, and performs perhaps the
best among all of the approximate methods tested. CCSDTQ-3 and CC4
further increase the cost to between 6 and 10\% of CCSDTQ. CCSDTQ-3
is somewhat inconsistent in predicting both geometric and energetic
quantities, and due to its increased cost relative to CCSDTQ-1b, the
former seems better justified as a standard iterative approximation.
CC4 also performs consistently well, but again entails a cost increase
over CCSDTQ-1b. Depending on whether an iterative or non-iterative
approximation is appropriate for the system in question, a roughly
one to two order-of-magnitude reduction in cost relative to full CCSDTQ
can be expected, with very nearly the same level of accuracy.

\section{Conclusions}

The application of approximate quadruples methods to the equilibrium
geometries and harmonic vibrational frequencies of the dimethylcarbene
to propene isomerization pathway system shows a uniformly good comparison
to the full CCSDTQ results, perhaps with the exception of the CCSDTQ-1a
method. Since this method alone lacks a direct coupling from the quadruples
amplitudes back to the triples, this seems the most likely explanation
for this poor behavior. Even for DMC, which might be expected to exhibit
mild multi-reference character due to the biradical carbon center,
the performance of iterative and non-iterative approximate quadruples
method alike remains very good. CCSD(T) does indeed show an increase
in error in this case, which can be traced almost exclusively to the
interaction between the two \emph{gauche} hydrogens which, presumably,
has important many-body contributions. For these and, we predict,
other ``well-behaved'' molecular systems, the CCSDT(Q) method, or
one of the two variants developed to rigorously treat the case of
a non-Hartree-Fock reference state, seem good choices that maximize
the accuracy-to-computational cost ratio. In particular, based on
the effect of non-bonded interactions in DMC, we expect that the inclusion
of quadruples effects in the calculation of equilibrium geometries
and (harmonic) vibrational frequencies to be important for dispersion-bound
complexes and strained molecules.

The application to the simplest Criegee Intermediate is much more
mixed, and clear evidence of multi-reference character is seen in
e.g. the magnitude of the converged coupled cluster amplitudes. In
this case, all methods exhibit a degradation of error with respect
to CCSDTQ, but in particular, the CCSDT(Q) method (and to a lesser
extent its A and B variants) as well as CCSDTQ-3 show a distinct worsening
in accuracy relative to CCSD(T), CCSDT, and the other approximate
quadruples methods. CCSDT(Q) performs especially poorly, and in fact
worsens the agreement with full CCSDTQ compared to CCSD(T) in every
regard except the total energy. Since the vast majority of previous
studies so far have focused on the energy alone in evaluating CCSDT(Q)
and other approximate quadruples methods, it seems important to consider
derived properties in order to gain a full picture of the intrinsic
accuracy of such methods. The results presented here suggest caution
when applying CCSDT(Q) to moderately multi-reference systems, although
further benchmarking is necessary to fully delineate the realm of
reasonable applicability. Luckily, the A and B variant show a distinct
improvement over ``plain'' CCSDT(Q) even in the restricted Hartree-Fock
case and serve as convenient substitutes. CCSDT(Q)/A can be implemented
at essentially no additional cost compared to CCSDT(Q), although the
current implementation in CFOUR is not optimal. Likewise CCSDTQ-1b
is seen to perform well in all circumstances tested and could in some
cases be justified as a higher-cost alternative.

In sum, these results show that the application of post-CCSDT methods
to the problem of equilibrium geometries, harmonic frequencies, and
potentially to other properties as well is both feasible and potentially
worthwhile when very high accuracy is necessary. Because of the computational
cost of the CCSDTQ benchmark calculations, only molecules with three
first-row atoms were investigated here. However, molecules with as
many as six first-row atoms should be accessible at the double-$\zeta$
level using the analytic gradient theory for approximate quadruples
methods.
\begin{acknowledgement}
This work was supported by a generous start-up grant from SMU, and
all calculations were performed on the ManeFrame II system at the
SMU Center for Scientific Computation. I would like to specially thank
Dr. John Stanton for his unrelenting encouragement, which was instrumental
in bringing this work to completion.
\end{acknowledgement}
\begin{suppinfo}
A supporting information file is available which contains individual
errors for each geometrical and vibrational parameter as well as the
CCSDTQ benchmark values.
\end{suppinfo}
\bibliography{biblio}

\providecommand{\latin}[1]{#1}
\makeatletter
\providecommand{\doi}
  {\begingroup\let\do\@makeother\dospecials
  \catcode`\{=1 \catcode`\}=2 \doi@aux}
\providecommand{\doi@aux}[1]{\endgroup\texttt{#1}}
\makeatother
\providecommand*\mcitethebibliography{\thebibliography}
\csname @ifundefined\endcsname{endmcitethebibliography}
  {\let\endmcitethebibliography\endthebibliography}{}
\begin{mcitethebibliography}{65}
\providecommand*\natexlab[1]{#1}
\providecommand*\mciteSetBstSublistMode[1]{}
\providecommand*\mciteSetBstMaxWidthForm[2]{}
\providecommand*\mciteBstWouldAddEndPuncttrue
  {\def\EndOfBibitem{\unskip.}}
\providecommand*\mciteBstWouldAddEndPunctfalse
  {\let\EndOfBibitem\relax}
\providecommand*\mciteSetBstMidEndSepPunct[3]{}
\providecommand*\mciteSetBstSublistLabelBeginEnd[3]{}
\providecommand*\EndOfBibitem{}
\mciteSetBstSublistMode{f}
\mciteSetBstMaxWidthForm{subitem}{(\alph{mcitesubitemcount})}
\mciteSetBstSublistLabelBeginEnd
  {\mcitemaxwidthsubitemform\space}
  {\relax}
  {\relax}

\bibitem[Martin(2005)]{martinChapterComputationalThermochemistry2005}
Martin,~J. M.~L. \emph{Annual {{Reports}} in {{Computational Chemistry}}};
  {Elsevier}, 2005; Vol.~1; pp 31--43\relax
\mciteBstWouldAddEndPuncttrue
\mciteSetBstMidEndSepPunct{\mcitedefaultmidpunct}
{\mcitedefaultendpunct}{\mcitedefaultseppunct}\relax
\EndOfBibitem
\bibitem[Feller \latin{et~al.}(2008)Feller, Peterson, and
  Dixon]{fellerSurveyFactorsContributing2008}
Feller,~D.; Peterson,~K.~A.; Dixon,~D.~A. A Survey of Factors Contributing to
  Accurate Theoretical Predictions of Atomization Energies and Molecular
  Structures. \emph{J. Chem. Phys.} \textbf{2008}, \emph{129}, 204105\relax
\mciteBstWouldAddEndPuncttrue
\mciteSetBstMidEndSepPunct{\mcitedefaultmidpunct}
{\mcitedefaultendpunct}{\mcitedefaultseppunct}\relax
\EndOfBibitem
\bibitem[Haunschild and Klopper(2012)Haunschild, and
  Klopper]{haunschildNewAccurateReference2012}
Haunschild,~R.; Klopper,~W. New Accurate Reference Energies for the {{G2}}/97
  Test Set. \emph{J. Chem. Phys.} \textbf{2012}, \emph{136}, 164102\relax
\mciteBstWouldAddEndPuncttrue
\mciteSetBstMidEndSepPunct{\mcitedefaultmidpunct}
{\mcitedefaultendpunct}{\mcitedefaultseppunct}\relax
\EndOfBibitem
\bibitem[Dixon \latin{et~al.}(2012)Dixon, Feller, and
  Peterson]{dixonChapterOnePractical2012}
Dixon,~D.~A.; Feller,~D.; Peterson,~K.~A. In \emph{Annual {{Reports}} in
  {{Computational Chemistry}}}; Wheeler,~R.~A., Ed.; Annual {{Reports}} in
  {{Computational Chemistry}}; {Elsevier}, 2012; Vol.~8; pp 1--28\relax
\mciteBstWouldAddEndPuncttrue
\mciteSetBstMidEndSepPunct{\mcitedefaultmidpunct}
{\mcitedefaultendpunct}{\mcitedefaultseppunct}\relax
\EndOfBibitem
\bibitem[Peterson \latin{et~al.}(2012)Peterson, Feller, and
  Dixon]{petersonChemicalAccuracyInitio2012}
Peterson,~K.~A.; Feller,~D.; Dixon,~D.~A. Chemical Accuracy in Ab Initio
  Thermochemistry and Spectroscopy: Current Strategies and Future Challenges.
  \emph{Theor Chem Acc} \textbf{2012}, \emph{131}, 1079\relax
\mciteBstWouldAddEndPuncttrue
\mciteSetBstMidEndSepPunct{\mcitedefaultmidpunct}
{\mcitedefaultendpunct}{\mcitedefaultseppunct}\relax
\EndOfBibitem
\bibitem[Dixon \latin{et~al.}(2001)Dixon, Feller, and
  Peterson]{dixonHeatsFormationIonization2001}
Dixon,~D.~A.; Feller,~D.; Peterson,~K.~A. Heats of Formation and Ionization
  Energies of {{NHx}}, X=0\textendash{}3. \emph{J. Chem. Phys.} \textbf{2001},
  \emph{115}, 2576--2581\relax
\mciteBstWouldAddEndPuncttrue
\mciteSetBstMidEndSepPunct{\mcitedefaultmidpunct}
{\mcitedefaultendpunct}{\mcitedefaultseppunct}\relax
\EndOfBibitem
\bibitem[Cs{\'a}sz{\'a}r \latin{et~al.}(2003)Cs{\'a}sz{\'a}r, Leininger, and
  Szalay]{csaszarStandardEnthalpyFormation2003}
Cs{\'a}sz{\'a}r,~A.~G.; Leininger,~M.~L.; Szalay,~V. The Standard Enthalpy of
  Formation of {{CH2}}. \emph{J. Chem. Phys.} \textbf{2003}, \emph{118},
  10631--10642\relax
\mciteBstWouldAddEndPuncttrue
\mciteSetBstMidEndSepPunct{\mcitedefaultmidpunct}
{\mcitedefaultendpunct}{\mcitedefaultseppunct}\relax
\EndOfBibitem
\bibitem[Feller \latin{et~al.}(2017)Feller, Bross, and
  Ruscic]{fellerEnthalpyFormationN2H42017}
Feller,~D.; Bross,~D.~H.; Ruscic,~B. Enthalpy of {{Formation}} of {{N2H4}}
  ({{Hydrazine}}) {{Revisited}}. \emph{J. Phys. Chem. A} \textbf{2017},
  \emph{121}, 6187--6198\relax
\mciteBstWouldAddEndPuncttrue
\mciteSetBstMidEndSepPunct{\mcitedefaultmidpunct}
{\mcitedefaultendpunct}{\mcitedefaultseppunct}\relax
\EndOfBibitem
\bibitem[Karton \latin{et~al.}(2009)Karton, Tarnopolsky, and
  Martin]{kartonAtomizationEnergiesCarbon2009}
Karton,~A.; Tarnopolsky,~A.; Martin,~J. M.~L. Atomization Energies of the
  Carbon Clusters {{C}} n (n = 2-10) Revisited by Means of {{W4}} Theory as
  Well as Density Functional, {{Gn}}, and {{CBS}} Methods. \emph{Mol. Phys.}
  \textbf{2009}, \emph{107}, 977--990, \_eprint:
  https://doi.org/10.1080/00268970802708959\relax
\mciteBstWouldAddEndPuncttrue
\mciteSetBstMidEndSepPunct{\mcitedefaultmidpunct}
{\mcitedefaultendpunct}{\mcitedefaultseppunct}\relax
\EndOfBibitem
\bibitem[Ruden \latin{et~al.}(2003)Ruden, Helgaker, J{\o}rgensen, and
  Olsen]{rudenCoupledclusterConnectedquadruplesCorrections2003}
Ruden,~T.~A.; Helgaker,~T.; J{\o}rgensen,~P.; Olsen,~J. Coupled-Cluster
  Connected-Quadruples Corrections to Atomization Energies. \emph{Chemical
  Physics Letters} \textbf{2003}, \emph{371}, 62--67\relax
\mciteBstWouldAddEndPuncttrue
\mciteSetBstMidEndSepPunct{\mcitedefaultmidpunct}
{\mcitedefaultendpunct}{\mcitedefaultseppunct}\relax
\EndOfBibitem
\bibitem[Werner \latin{et~al.}(2008)Werner, K{\'a}llay, and
  Gauss]{wernerBarrierHeightH22008}
Werner,~H.-J.; K{\'a}llay,~M.; Gauss,~J. The Barrier Height of the {{F}}+{{H2}}
  Reaction Revisited: {{Coupled}}-Cluster and Multireference
  Configuration-Interaction Benchmark Calculations. \emph{J. Chem. Phys.}
  \textbf{2008}, \emph{128}, 034305\relax
\mciteBstWouldAddEndPuncttrue
\mciteSetBstMidEndSepPunct{\mcitedefaultmidpunct}
{\mcitedefaultendpunct}{\mcitedefaultseppunct}\relax
\EndOfBibitem
\bibitem[Hopkins and Tschumper(2004)Hopkins, and
  Tschumper]{hopkinsInitioStudiesInteractions2004}
Hopkins,~B.~W.; Tschumper,~G.~S. Ab {{Initio Studies}} of
  {$\Pi\cdot\cdot\cdot\pi$} {{Interactions}}:\, {{The Effects}} of {{Quadruple
  Excitations}}\textdagger{}. \emph{J. Phys. Chem. A} \textbf{2004},
  \emph{108}, 2941--2948\relax
\mciteBstWouldAddEndPuncttrue
\mciteSetBstMidEndSepPunct{\mcitedefaultmidpunct}
{\mcitedefaultendpunct}{\mcitedefaultseppunct}\relax
\EndOfBibitem
\bibitem[Tajti \latin{et~al.}(2004)Tajti, Szalay, Cs{\'a}sz{\'a}r, K{\'a}llay,
  Gauss, Valeev, Flowers, V{\'a}zquez, and Stanton]{tajtiHEATHighAccuracy2004}
Tajti,~A.; Szalay,~P.~G.; Cs{\'a}sz{\'a}r,~A.~G.; K{\'a}llay,~M.; Gauss,~J.;
  Valeev,~E.~F.; Flowers,~B.~A.; V{\'a}zquez,~J.; Stanton,~J.~F. {{HEAT}}:
  {{High}} Accuracy Extrapolated Ab Initio Thermochemistry. \emph{J. Chem.
  Phys.} \textbf{2004}, \emph{121}, 11599--11613\relax
\mciteBstWouldAddEndPuncttrue
\mciteSetBstMidEndSepPunct{\mcitedefaultmidpunct}
{\mcitedefaultendpunct}{\mcitedefaultseppunct}\relax
\EndOfBibitem
\bibitem[Bomble \latin{et~al.}(2006)Bomble, V{\'a}zquez, K{\'a}llay, Michauk,
  Szalay, Cs{\'a}sz{\'a}r, Gauss, and
  Stanton]{bombleHighaccuracyExtrapolatedInitio2006}
Bomble,~Y.~J.; V{\'a}zquez,~J.; K{\'a}llay,~M.; Michauk,~C.; Szalay,~P.~G.;
  Cs{\'a}sz{\'a}r,~A.~G.; Gauss,~J.; Stanton,~J.~F. High-Accuracy Extrapolated
  Ab Initio Thermochemistry. {{II}}. {{Minor}} Improvements to the Protocol and
  a Vital Simplification. \emph{J. Chem. Phys.} \textbf{2006}, \emph{125},
  064108\relax
\mciteBstWouldAddEndPuncttrue
\mciteSetBstMidEndSepPunct{\mcitedefaultmidpunct}
{\mcitedefaultendpunct}{\mcitedefaultseppunct}\relax
\EndOfBibitem
\bibitem[Harding \latin{et~al.}(2008)Harding, V{\'a}zquez, Ruscic, Wilson,
  Gauss, and Stanton]{hardingHighaccuracyExtrapolatedInitio2008}
Harding,~M.~E.; V{\'a}zquez,~J.; Ruscic,~B.; Wilson,~A.~K.; Gauss,~J.;
  Stanton,~J.~F. High-Accuracy Extrapolated Ab Initio Thermochemistry. {{III}}.
  {{Additional}} Improvements and Overview. \emph{J. Chem. Phys.}
  \textbf{2008}, \emph{128}, 114111\relax
\mciteBstWouldAddEndPuncttrue
\mciteSetBstMidEndSepPunct{\mcitedefaultmidpunct}
{\mcitedefaultendpunct}{\mcitedefaultseppunct}\relax
\EndOfBibitem
\bibitem[Thorpe \latin{et~al.}(2019)Thorpe, Lopez, Nguyen, Baraban, Bross,
  Ruscic, and Stanton]{thorpeHighaccuracyExtrapolatedInitio2019}
Thorpe,~J.~H.; Lopez,~C.~A.; Nguyen,~T.~L.; Baraban,~J.~H.; Bross,~D.~H.;
  Ruscic,~B.; Stanton,~J.~F. High-Accuracy Extrapolated Ab Initio
  Thermochemistry. {{IV}}. {{A}} Modified Recipe for Computational Efficiency.
  \emph{J. Chem. Phys.} \textbf{2019}, \emph{150}, 224102\relax
\mciteBstWouldAddEndPuncttrue
\mciteSetBstMidEndSepPunct{\mcitedefaultmidpunct}
{\mcitedefaultendpunct}{\mcitedefaultseppunct}\relax
\EndOfBibitem
\bibitem[Martin and de~Oliveira(1999)Martin, and
  de~Oliveira]{martinStandardMethodsBenchmark1999}
Martin,~J. M.~L.; de~Oliveira,~G. Towards Standard Methods for Benchmark
  Quality Ab Initio Thermochemistry\textemdash{{W1}} and {{W2}} Theory.
  \emph{J. Chem. Phys.} \textbf{1999}, \emph{111}, 1843--1856\relax
\mciteBstWouldAddEndPuncttrue
\mciteSetBstMidEndSepPunct{\mcitedefaultmidpunct}
{\mcitedefaultendpunct}{\mcitedefaultseppunct}\relax
\EndOfBibitem
\bibitem[Boese \latin{et~al.}(2004)Boese, Oren, Atasoylu, Martin, K{\'a}llay,
  and Gauss]{boeseW3TheoryRobust2004}
Boese,~A.~D.; Oren,~M.; Atasoylu,~O.; Martin,~J. M.~L.; K{\'a}llay,~M.;
  Gauss,~J. W3 Theory: {{Robust}} Computational Thermochemistry in the
  {{kJ}}/Mol Accuracy Range. \emph{J. Chem. Phys.} \textbf{2004}, \emph{120},
  4129--4141\relax
\mciteBstWouldAddEndPuncttrue
\mciteSetBstMidEndSepPunct{\mcitedefaultmidpunct}
{\mcitedefaultendpunct}{\mcitedefaultseppunct}\relax
\EndOfBibitem
\bibitem[Karton \latin{et~al.}(2006)Karton, Rabinovich, Martin, and
  Ruscic]{kartonW4TheoryComputational2006}
Karton,~A.; Rabinovich,~E.; Martin,~J. M.~L.; Ruscic,~B. W4 Theory for
  Computational Thermochemistry: {{In}} Pursuit of Confident Sub-{{kJ}}/Mol
  Predictions. \emph{J. Chem. Phys.} \textbf{2006}, \emph{125}, 144108\relax
\mciteBstWouldAddEndPuncttrue
\mciteSetBstMidEndSepPunct{\mcitedefaultmidpunct}
{\mcitedefaultendpunct}{\mcitedefaultseppunct}\relax
\EndOfBibitem
\bibitem[Klippenstein \latin{et~al.}(2017)Klippenstein, Harding, and
  Ruscic]{klippensteinInitioComputationsActive2017}
Klippenstein,~S.~J.; Harding,~L.~B.; Ruscic,~B. Ab {{Initio Computations}} and
  {{Active Thermochemical Tables Hand}} in {{Hand}}: {{Heats}} of {{Formation}}
  of {{Core Combustion Species}}. \emph{J. Phys. Chem. A} \textbf{2017},
  \emph{121}, 6580--6602\relax
\mciteBstWouldAddEndPuncttrue
\mciteSetBstMidEndSepPunct{\mcitedefaultmidpunct}
{\mcitedefaultendpunct}{\mcitedefaultseppunct}\relax
\EndOfBibitem
\bibitem[Raghavachari \latin{et~al.}(1989)Raghavachari, Trucks, Pople, and
  {Head-Gordon}]{raghavachariFifthorderPerturbationComparison1989a}
Raghavachari,~K.; Trucks,~G.~W.; Pople,~J.~A.; {Head-Gordon},~M. A Fifth-Order
  Perturbation Comparison of Electron Correlation Theories. \emph{Chemical
  Physics Letters} \textbf{1989}, \emph{157}, 479--483\relax
\mciteBstWouldAddEndPuncttrue
\mciteSetBstMidEndSepPunct{\mcitedefaultmidpunct}
{\mcitedefaultendpunct}{\mcitedefaultseppunct}\relax
\EndOfBibitem
\bibitem[Noga and Bartlett(1987)Noga, and Bartlett]{nogaFullCCSDTModel1987a}
Noga,~J.; Bartlett,~R.~J. The Full {{CCSDT}} Model for Molecular Electronic
  Structure. \emph{J. Chem. Phys.} \textbf{1987}, \emph{86}, 7041\relax
\mciteBstWouldAddEndPuncttrue
\mciteSetBstMidEndSepPunct{\mcitedefaultmidpunct}
{\mcitedefaultendpunct}{\mcitedefaultseppunct}\relax
\EndOfBibitem
\bibitem[Bomble \latin{et~al.}(2005)Bomble, Stanton, K{\'a}llay, and
  Gauss]{bombleCoupledclusterMethodsIncluding2005}
Bomble,~Y.~J.; Stanton,~J.~F.; K{\'a}llay,~M.; Gauss,~J. Coupled-Cluster
  Methods Including Noniterative Corrections for Quadruple Excitations.
  \emph{Journal of Chemical Physics} \textbf{2005}, \emph{123}, 4101\relax
\mciteBstWouldAddEndPuncttrue
\mciteSetBstMidEndSepPunct{\mcitedefaultmidpunct}
{\mcitedefaultendpunct}{\mcitedefaultseppunct}\relax
\EndOfBibitem
\bibitem[Kucharski and Bartlett(1991)Kucharski, and
  Bartlett]{kucharskiRecursiveIntermediateFactorization1991}
Kucharski,~S.~A.; Bartlett,~R.~J. Recursive Intermediate Factorization and
  Complete Computational Linearization of the Coupled-Cluster Single, Double,
  Triple, and Quadruple Excitation Equations. \emph{Theor. Chem. Acc. Theory
  Comput. Model. Theor. Chim. Acta} \textbf{1991}, \emph{80}, 387--405\relax
\mciteBstWouldAddEndPuncttrue
\mciteSetBstMidEndSepPunct{\mcitedefaultmidpunct}
{\mcitedefaultendpunct}{\mcitedefaultseppunct}\relax
\EndOfBibitem
\bibitem[Oliphant and Adamowicz(1991)Oliphant, and
  Adamowicz]{oliphantCoupledClusterMethod1991}
Oliphant,~N.; Adamowicz,~L. Coupled-cluster Method Truncated at Quadruples.
  \emph{J. Chem. Phys.} \textbf{1991}, \emph{95}, 6645--6651\relax
\mciteBstWouldAddEndPuncttrue
\mciteSetBstMidEndSepPunct{\mcitedefaultmidpunct}
{\mcitedefaultendpunct}{\mcitedefaultseppunct}\relax
\EndOfBibitem
\bibitem[Kucharski and Bartlett(1992)Kucharski, and
  Bartlett]{kucharskiCoupledClusterSingle1992}
Kucharski,~S.~A.; Bartlett,~R.~J. The Coupled-cluster Single, Double, Triple,
  and Quadruple Excitation Method. \emph{J. Chem. Phys.} \textbf{1992},
  \emph{97}, 4282--4288\relax
\mciteBstWouldAddEndPuncttrue
\mciteSetBstMidEndSepPunct{\mcitedefaultmidpunct}
{\mcitedefaultendpunct}{\mcitedefaultseppunct}\relax
\EndOfBibitem
\bibitem[Morgan \latin{et~al.}(2018)Morgan, Matthews, Ringholm, Agarwal, Gong,
  Ruud, Allen, Stanton, and Schaefer]{morganGeometricEnergyDerivatives2018}
Morgan,~W.~J.; Matthews,~D.~A.; Ringholm,~M.; Agarwal,~J.; Gong,~J.~Z.;
  Ruud,~K.; Allen,~W.~D.; Stanton,~J.~F.; Schaefer,~H.~F. Geometric {{Energy
  Derivatives}} at the {{Complete Basis Set Limit}}: {{Application}} to the
  {{Equilibrium Structure}} and {{Molecular Force Field}} of {{Formaldehyde}}.
  \emph{J. Chem. Theory Comput.} \textbf{2018}, \emph{14}, 1333--1350\relax
\mciteBstWouldAddEndPuncttrue
\mciteSetBstMidEndSepPunct{\mcitedefaultmidpunct}
{\mcitedefaultendpunct}{\mcitedefaultseppunct}\relax
\EndOfBibitem
\bibitem[Puzzarini \latin{et~al.}(2008)Puzzarini, Heckert, and
  Gauss]{puzzariniAccuracyRotationalConstants2008}
Puzzarini,~C.; Heckert,~M.; Gauss,~J. The Accuracy of Rotational Constants
  Predicted by High-Level Quantum-Chemical Calculations. {{I}}. Molecules
  Containing First-Row Atoms. \emph{J. Chem. Phys.} \textbf{2008}, \emph{128},
  194108\relax
\mciteBstWouldAddEndPuncttrue
\mciteSetBstMidEndSepPunct{\mcitedefaultmidpunct}
{\mcitedefaultendpunct}{\mcitedefaultseppunct}\relax
\EndOfBibitem
\bibitem[Heckert \latin{et~al.}(2005)Heckert, K{\'a}llay, and
  Gauss]{heckertMolecularEquilibriumGeometries2005a}
Heckert,~M.; K{\'a}llay,~M.; Gauss,~J. Molecular Equilibrium Geometries Based
  on Coupled-Cluster Calculations Including Quadruple Excitations. \emph{Mol.
  Phys.} \textbf{2005}, \emph{103}, 2109--2115, \_eprint:
  https://doi.org/10.1080/00268970500083416\relax
\mciteBstWouldAddEndPuncttrue
\mciteSetBstMidEndSepPunct{\mcitedefaultmidpunct}
{\mcitedefaultendpunct}{\mcitedefaultseppunct}\relax
\EndOfBibitem
\bibitem[Heckert \latin{et~al.}(2006)Heckert, K{\'a}llay, Tew, Klopper, and
  Gauss]{heckertBasissetExtrapolationTechniques2006}
Heckert,~M.; K{\'a}llay,~M.; Tew,~D.~P.; Klopper,~W.; Gauss,~J. Basis-Set
  Extrapolation Techniques for the Accurate Calculation of Molecular
  Equilibrium Geometries Using Coupled-Cluster Theory. \emph{J. Chem. Phys.}
  \textbf{2006}, \emph{125}, 044108\relax
\mciteBstWouldAddEndPuncttrue
\mciteSetBstMidEndSepPunct{\mcitedefaultmidpunct}
{\mcitedefaultendpunct}{\mcitedefaultseppunct}\relax
\EndOfBibitem
\bibitem[Ruden \latin{et~al.}(2004)Ruden, Helgaker, J{\o}rgensen, and
  Olsen]{rudenCoupledclusterConnectedQuadruples2004}
Ruden,~T.~A.; Helgaker,~T.; J{\o}rgensen,~P.; Olsen,~J. Coupled-Cluster
  Connected Quadruples and Quintuples Corrections to the Harmonic Vibrational
  Frequencies and Equilibrium Bond Distances of {{HF}}, {{N2}}, {{F2}}, and
  {{CO}}. \emph{J. Chem. Phys.} \textbf{2004}, \emph{121}, 5874--5884\relax
\mciteBstWouldAddEndPuncttrue
\mciteSetBstMidEndSepPunct{\mcitedefaultmidpunct}
{\mcitedefaultendpunct}{\mcitedefaultseppunct}\relax
\EndOfBibitem
\bibitem[Handy and Schaefer(1984)Handy, and
  Schaefer]{handyEvaluationAnalyticEnergy1984}
Handy,~N.~C.; Schaefer,~H.~F. On the Evaluation of Analytic Energy Derivatives
  for Correlated Wave Functions. \emph{J. Chem. Phys.} \textbf{1984},
  \emph{81}, 5031--5033\relax
\mciteBstWouldAddEndPuncttrue
\mciteSetBstMidEndSepPunct{\mcitedefaultmidpunct}
{\mcitedefaultendpunct}{\mcitedefaultseppunct}\relax
\EndOfBibitem
\bibitem[K{\'a}llay and Gauss(2008)K{\'a}llay, and
  Gauss]{kallayApproximateTreatmentHigher2008}
K{\'a}llay,~M.; Gauss,~J. Approximate Treatment of Higher Excitations in
  Coupled-Cluster Theory. {{II}}. {{Extension}} to General Single-Determinant
  Reference Functions and Improved Approaches for the Canonical
  {{Hartree}}\textendash{{Fock}} Case. \emph{J. Chem. Phys.} \textbf{2008},
  \emph{129}, 144101\relax
\mciteBstWouldAddEndPuncttrue
\mciteSetBstMidEndSepPunct{\mcitedefaultmidpunct}
{\mcitedefaultendpunct}{\mcitedefaultseppunct}\relax
\EndOfBibitem
\bibitem[K{\'a}llay and Gauss(2005)K{\'a}llay, and
  Gauss]{kallayApproximateTreatmentHigher2005}
K{\'a}llay,~M.; Gauss,~J. Approximate Treatment of Higher Excitations in
  Coupled-Cluster Theory. \emph{J. Chem. Phys.} \textbf{2005}, \emph{123},
  214105--214105--13\relax
\mciteBstWouldAddEndPuncttrue
\mciteSetBstMidEndSepPunct{\mcitedefaultmidpunct}
{\mcitedefaultendpunct}{\mcitedefaultseppunct}\relax
\EndOfBibitem
\bibitem[Gauss and Stanton(2002)Gauss, and
  Stanton]{gaussAnalyticGradientsCoupledcluster2002}
Gauss,~J.; Stanton,~J.~F. Analytic Gradients for the Coupled-Cluster Singles,
  Doubles, and Triples ({{CCSDT}}) Model. \emph{J. Chem. Phys.} \textbf{2002},
  \emph{116}, 1773\relax
\mciteBstWouldAddEndPuncttrue
\mciteSetBstMidEndSepPunct{\mcitedefaultmidpunct}
{\mcitedefaultendpunct}{\mcitedefaultseppunct}\relax
\EndOfBibitem
\bibitem[K{\'a}llay \latin{et~al.}(2003)K{\'a}llay, Gauss, and
  Szalay]{kallayAnalyticFirstDerivatives2003}
K{\'a}llay,~M.; Gauss,~J.; Szalay,~P.~G. Analytic First Derivatives for General
  Coupled-Cluster and Configuration Interaction Models. \emph{J. Chem. Phys.}
  \textbf{2003}, \emph{119}, 2991--3004\relax
\mciteBstWouldAddEndPuncttrue
\mciteSetBstMidEndSepPunct{\mcitedefaultmidpunct}
{\mcitedefaultendpunct}{\mcitedefaultseppunct}\relax
\EndOfBibitem
\bibitem[Scuseria and Schaefer(1988)Scuseria, and
  Schaefer]{scuseriaAnalyticEvaluationEnergy1988}
Scuseria,~G.~E.; Schaefer,~H.~F. Analytic Evaluation of Energy Gradients for
  the Single, Double and Linearized Triple Excitation Coupled-Cluster
  {{CCSDT}}-1 Wavefunction: {{Theory}} and Applications. \emph{Chem. Phys.
  Lett.} \textbf{1988}, \emph{146}, 23--31\relax
\mciteBstWouldAddEndPuncttrue
\mciteSetBstMidEndSepPunct{\mcitedefaultmidpunct}
{\mcitedefaultendpunct}{\mcitedefaultseppunct}\relax
\EndOfBibitem
\bibitem[Gauss and Stanton(2000)Gauss, and
  Stanton]{gaussAnalyticFirstSecond2000}
Gauss,~J.; Stanton,~J.~F. Analytic First and Second Derivatives for the
  {{CCSDT}}-n (N=1\textendash{}3) Models: A First Step towards the Efficient
  Calculation of {{CCSDT}} Properties. \emph{Phys. Chem. Chem. Phys.}
  \textbf{2000}, \emph{2}, 2047--2060\relax
\mciteBstWouldAddEndPuncttrue
\mciteSetBstMidEndSepPunct{\mcitedefaultmidpunct}
{\mcitedefaultendpunct}{\mcitedefaultseppunct}\relax
\EndOfBibitem
\bibitem[Scuseria(1991)]{scuseriaAnalyticEvaluationEnergy1991}
Scuseria,~G.~E. Analytic Evaluation of Energy Gradients for the Singles and
  Doubles Coupled Cluster Method Including Perturbative Triple Excitations:
  {{Theory}} and Applications to {{FOOF}} and {{Cr2}}. \emph{J. Chem. Phys.}
  \textbf{1991}, \emph{94}, 442\relax
\mciteBstWouldAddEndPuncttrue
\mciteSetBstMidEndSepPunct{\mcitedefaultmidpunct}
{\mcitedefaultendpunct}{\mcitedefaultseppunct}\relax
\EndOfBibitem
\bibitem[Watts \latin{et~al.}(1992)Watts, Gauss, and
  Bartlett]{wattsOpenshellAnalyticalEnergy1992}
Watts,~J.~D.; Gauss,~J.; Bartlett,~R.~J. Open-Shell Analytical Energy Gradients
  for Triple Excitation Many-Body, Coupled-Cluster Methods: {{MBPT}}(4),
  {{CCSD}}+{{T}}({{CCSD}}), {{CCSD}}({{T}}),and {{QCISD}}({{T}}).
  \emph{Chemical Physics Letters} \textbf{1992}, \emph{200}, 1--7\relax
\mciteBstWouldAddEndPuncttrue
\mciteSetBstMidEndSepPunct{\mcitedefaultmidpunct}
{\mcitedefaultendpunct}{\mcitedefaultseppunct}\relax
\EndOfBibitem
\bibitem[Scheiner \latin{et~al.}(1987)Scheiner, Scuseria, Rice, Lee, and
  Schaefer]{scheinerAnalyticEvaluationEnergy1987}
Scheiner,~A.~C.; Scuseria,~G.~E.; Rice,~J.~E.; Lee,~T.~J.; Schaefer,~H.~F.
  Analytic Evaluation of Energy Gradients for the Single and Double Excitation
  Coupled Cluster ({{CCSD}}) Wave Function: {{Theory}} and Application.
  \emph{J. Chem. Phys.} \textbf{1987}, \emph{87}, 5361\relax
\mciteBstWouldAddEndPuncttrue
\mciteSetBstMidEndSepPunct{\mcitedefaultmidpunct}
{\mcitedefaultendpunct}{\mcitedefaultseppunct}\relax
\EndOfBibitem
\bibitem[Rendell and Lee(1991)Rendell, and
  Lee]{rendellEfficientFormulationImplementation1991}
Rendell,~A.~P.; Lee,~T.~J. An Efficient Formulation and Implementation of the
  Analytic Energy Gradient Method to the Single and Double Excitation
  Coupled-Cluster Wave Function: {{Application}} to {{Cl2O2}}. \emph{J. Chem.
  Phys.} \textbf{1991}, \emph{94}, 6219\relax
\mciteBstWouldAddEndPuncttrue
\mciteSetBstMidEndSepPunct{\mcitedefaultmidpunct}
{\mcitedefaultendpunct}{\mcitedefaultseppunct}\relax
\EndOfBibitem
\bibitem[Gauss \latin{et~al.}(1991)Gauss, Stanton, and
  Bartlett]{gaussCoupledClusterOpen1991}
Gauss,~J.; Stanton,~J.~F.; Bartlett,~R.~J. Coupled-cluster Open-shell Analytic
  Gradients: {{Implementation}} of the Direct Product Decomposition Approach in
  Energy Gradient Calculations. \emph{The Journal of Chemical Physics}
  \textbf{1991}, \emph{95}, 2623--2638\relax
\mciteBstWouldAddEndPuncttrue
\mciteSetBstMidEndSepPunct{\mcitedefaultmidpunct}
{\mcitedefaultendpunct}{\mcitedefaultseppunct}\relax
\EndOfBibitem
\bibitem[{\v C}{\'i}{\v z}ek(1966)]{cizekCorrelationProblemAtomic1966}
{\v C}{\'i}{\v z}ek,~J. On the {{Correlation Problem}} in {{Atomic}} and
  {{Molecular Systems}}. {{Calculation}} of {{Wavefunction Components}} in
  {{Ursell}}-{{Type Expansion Using Quantum}}-{{Field Theoretical Methods}}.
  \emph{J. Chem. Phys.} \textbf{1966}, \emph{45}, 4256--4266\relax
\mciteBstWouldAddEndPuncttrue
\mciteSetBstMidEndSepPunct{\mcitedefaultmidpunct}
{\mcitedefaultendpunct}{\mcitedefaultseppunct}\relax
\EndOfBibitem
\bibitem[Shavitt and Bartlett(2009)Shavitt, and
  Bartlett]{shavittManyBodyMethodsChemistry2009}
Shavitt,~I.; Bartlett,~R.~J. \emph{Many-{{Body Methods}} in {{Chemistry}} and
  {{Physics}}: {{MBPT}} and {{Coupled}}-{{Cluster Theory}}}, 1st ed.;
  {Cambridge University Press}: {Cambridge ; New York}, 2009\relax
\mciteBstWouldAddEndPuncttrue
\mciteSetBstMidEndSepPunct{\mcitedefaultmidpunct}
{\mcitedefaultendpunct}{\mcitedefaultseppunct}\relax
\EndOfBibitem
\bibitem[Helgaker \latin{et~al.}(2013)Helgaker, Jorgensen, and
  Olsen]{helgakerMolecularElectronicStructureTheory2013}
Helgaker,~T.; Jorgensen,~P.; Olsen,~J. \emph{Molecular
  {{Electronic}}-{{Structure Theory}}}, 1st ed.; {Wiley}: {Chichester ; New
  York}, 2013\relax
\mciteBstWouldAddEndPuncttrue
\mciteSetBstMidEndSepPunct{\mcitedefaultmidpunct}
{\mcitedefaultendpunct}{\mcitedefaultseppunct}\relax
\EndOfBibitem
\bibitem[Rice \latin{et~al.}(1986)Rice, Amos, Handy, Lee, and
  Schaefer]{riceAnalyticConfigurationInteraction1986}
Rice,~J.~E.; Amos,~R.~D.; Handy,~N.~C.; Lee,~T.~J.; Schaefer,~H.~F. The
  Analytic Configuration Interaction Gradient Method: {{Application}} to the
  Cyclic and Open Isomers of the {{S3}} Molecule. \emph{J. Chem. Phys.}
  \textbf{1986}, \emph{85}, 963--968\relax
\mciteBstWouldAddEndPuncttrue
\mciteSetBstMidEndSepPunct{\mcitedefaultmidpunct}
{\mcitedefaultendpunct}{\mcitedefaultseppunct}\relax
\EndOfBibitem
\bibitem[Bartlett \latin{et~al.}(1990)Bartlett, Watts, Kucharski, and
  Noga]{bartlettNoniterativeFifthorderTriple1990}
Bartlett,~R.~J.; Watts,~J.~D.; Kucharski,~S.~A.; Noga,~J. Non-Iterative
  Fifth-Order Triple and Quadruple Excitation Energy Corrections in Correlated
  Methods. \emph{Chemical Physics Letters} \textbf{1990}, \emph{165},
  513--522\relax
\mciteBstWouldAddEndPuncttrue
\mciteSetBstMidEndSepPunct{\mcitedefaultmidpunct}
{\mcitedefaultendpunct}{\mcitedefaultseppunct}\relax
\EndOfBibitem
\bibitem[Stanton(1997)]{stantonWhyCCSDWorks1997}
Stanton,~J.~F. Why {{CCSD}}({{T}}) Works: A Different Perspective. \emph{Chem.
  Phys. Lett.} \textbf{1997}, \emph{281}, 130\relax
\mciteBstWouldAddEndPuncttrue
\mciteSetBstMidEndSepPunct{\mcitedefaultmidpunct}
{\mcitedefaultendpunct}{\mcitedefaultseppunct}\relax
\EndOfBibitem
\bibitem[Matthews and Stanton(2015)Matthews, and
  Stanton]{matthewsNonorthogonalSpinadaptationCoupled2015}
Matthews,~D.~A.; Stanton,~J.~F. Non-Orthogonal Spin-Adaptation of Coupled
  Cluster Methods: {{A}} New Implementation of Methods Including Quadruple
  Excitations. \emph{J. Chem. Phys.} \textbf{2015}, \emph{142}, 064108\relax
\mciteBstWouldAddEndPuncttrue
\mciteSetBstMidEndSepPunct{\mcitedefaultmidpunct}
{\mcitedefaultendpunct}{\mcitedefaultseppunct}\relax
\EndOfBibitem
\bibitem[Stanton \latin{et~al.}()Stanton, Gauss, Cheng, Harding, Matthews,
  Szalay, Auer, Bartlett, Benedikt, Berger, Bernholdt, Bomble, Christiansen,
  Engel, Faber, Heckert, Heun, Hilgenberg, Huber, Jagau, Jonsson, Juse{\l}ius,
  Kirsch, Klein, Lauderdale, Lipparini, Metzroth, M{\"u}ck, O'Neill, Price,
  Prochnow, Puzzarini, Ruud, Schiffmann, Schwalbach, Simmons, Stopkowicz,
  Tajti, V{\'a}zquez, Wang, and Watts]{stantonCFOURCoupledClusterTechniques}
Stanton,~J.~F.; Gauss,~J.; Cheng,~L.; Harding,~M.~E.; Matthews,~D.~A.;
  Szalay,~P.~G.; Auer,~A.~A.; Bartlett,~R.~J.; Benedikt,~U.; Berger,~C.;
  Bernholdt,~D.~E.; Bomble,~Y.~J.; Christiansen,~O.; Engel,~F.; Faber,~R.;
  Heckert,~M.; Heun,~O.; Hilgenberg,~M.; Huber,~C.; Jagau,~T.-C.; Jonsson,~D.;
  Juse{\l}ius,~J.; Kirsch,~T.; Klein,~K.; Lauderdale,~W.~J.; Lipparini,~F.;
  Metzroth,~T.; M{\"u}ck,~L.~A.; O'Neill,~D.~P.; Price,~D.~R.; Prochnow,~E.;
  Puzzarini,~C.; Ruud,~K.; Schiffmann,~F.; Schwalbach,~W.; Simmons,~C.;
  Stopkowicz,~S.; Tajti,~A.; V{\'a}zquez,~J.; Wang,~F.; Watts,~J.~D.
  \emph{{{CFOUR}}, {{Coupled}}-{{Cluster}} Techniques for {{Computational
  Chemistry}}, a Quantum-Chemical Program Package with the Integral Packages
  {{MOLECULE}} ({{J}}. {{Alml\"of}} and {{P}}. {{R}}. {{Taylor}}), {{PROPS}}
  ({{P}}. {{R}}. {{Taylor}}), {{ABACUS}} ({{T}}. {{Helgaker}}, {{H}}. {{J}}.
  {{Jensen}}, {{P}}. {{J\o{}rgensen}} and {{J}}. {{Olsen}}), and {{ECP}}
  Routines by {{A}}. {{V}}. {{Mitin}} and {{C}}. van {{W\"ullen}}};
  www.cfour.de\relax
\mciteBstWouldAddEndPuncttrue
\mciteSetBstMidEndSepPunct{\mcitedefaultmidpunct}
{\mcitedefaultendpunct}{\mcitedefaultseppunct}\relax
\EndOfBibitem
\bibitem[Matthews and Stanton(2019)Matthews, and
  Stanton]{matthewsChapter10Diagrams2019}
Matthews,~D.~A.; Stanton,~J.~F. In \emph{Mathematical {{Physics}} in
  {{Theoretical Chemistry}}}; Blinder,~S.~M., House,~J.~E., Eds.; Developments
  in {{Physical}} \& {{Theoretical Chemistry}}; {Elsevier}, 2019; pp
  327--375\relax
\mciteBstWouldAddEndPuncttrue
\mciteSetBstMidEndSepPunct{\mcitedefaultmidpunct}
{\mcitedefaultendpunct}{\mcitedefaultseppunct}\relax
\EndOfBibitem
\bibitem[Matthews(2018)]{matthewsHighPerformanceTensorContraction2018}
Matthews,~D. High-{{Performance Tensor Contraction}} without {{Transposition}}.
  \emph{SIAM J. Sci. Comput.} \textbf{2018}, \emph{40}, C1--C24\relax
\mciteBstWouldAddEndPuncttrue
\mciteSetBstMidEndSepPunct{\mcitedefaultmidpunct}
{\mcitedefaultendpunct}{\mcitedefaultseppunct}\relax
\EndOfBibitem
\bibitem[Matthews(2019)]{matthewsExtendingOptimisingDirect2019}
Matthews,~D.~A. On Extending and Optimising the Direct Product Decomposition.
  \emph{Mol. Phys.} \textbf{2019}, \emph{117}, 1325--1333\relax
\mciteBstWouldAddEndPuncttrue
\mciteSetBstMidEndSepPunct{\mcitedefaultmidpunct}
{\mcitedefaultendpunct}{\mcitedefaultseppunct}\relax
\EndOfBibitem
\bibitem[Matthews and Stanton(2015)Matthews, and
  Stanton]{matthewsAcceleratingConvergenceHigherorder2015}
Matthews,~D.~A.; Stanton,~J.~F. Accelerating the Convergence of Higher-Order
  Coupled Cluster Methods. \emph{J. Chem. Phys.} \textbf{2015}, \emph{143},
  204103\relax
\mciteBstWouldAddEndPuncttrue
\mciteSetBstMidEndSepPunct{\mcitedefaultmidpunct}
{\mcitedefaultendpunct}{\mcitedefaultseppunct}\relax
\EndOfBibitem
\bibitem[Matthews(2020)]{doi:10.1080/00268976.2020.1757774}
Matthews,~D. Accelerating the Convergence of Higher-Order Coupled Cluster
  Methods {{II}}: {{Coupled}} Cluster {{$\Lambda$}} Equations and Dynamic
  Damping. \emph{Mol. Phys.} \textbf{2020}, \relax
\mciteBstWouldAddEndPunctfalse
\mciteSetBstMidEndSepPunct{\mcitedefaultmidpunct}
{}{\mcitedefaultseppunct}\relax
\EndOfBibitem
\bibitem[K{\'a}llay \latin{et~al.}(2020)K{\'a}llay, Nagy, Mester, Rolik, Samu,
  Csontos, Cs{\'o}ka, Szab{\'o}, {Gyevi-Nagy}, H{\'e}gely, Ladj{\'a}nszki,
  Szegedy, Lad{\'o}czki, Petrov, Farkas, Mezei, and
  Ganyecz]{kallayMRCCProgramSystem2020}
K{\'a}llay,~M.; Nagy,~P.~R.; Mester,~D.; Rolik,~Z.; Samu,~G.; Csontos,~J.;
  Cs{\'o}ka,~J.; Szab{\'o},~P.~B.; {Gyevi-Nagy},~L.; H{\'e}gely,~B.;
  Ladj{\'a}nszki,~I.; Szegedy,~L.; Lad{\'o}czki,~B.; Petrov,~K.; Farkas,~M.;
  Mezei,~P.~D.; Ganyecz,~{\'A}. The {{MRCC}} Program System: {{Accurate}}
  Quantum Chemistry from Water to Proteins. \emph{J. Chem. Phys.}
  \textbf{2020}, \emph{ESS2020}, 074107\relax
\mciteBstWouldAddEndPuncttrue
\mciteSetBstMidEndSepPunct{\mcitedefaultmidpunct}
{\mcitedefaultendpunct}{\mcitedefaultseppunct}\relax
\EndOfBibitem
\bibitem[Criegee and Wenner(1949)Criegee, and
  Wenner]{criegeeOzonisierung10Oktalins1949}
Criegee,~R.; Wenner,~G. Die {{Ozonisierung}} Des 9,10-{{Oktalins}}.
  \emph{Justus Liebigs Annalen der Chemie} \textbf{1949}, \emph{564},
  9--15\relax
\mciteBstWouldAddEndPuncttrue
\mciteSetBstMidEndSepPunct{\mcitedefaultmidpunct}
{\mcitedefaultendpunct}{\mcitedefaultseppunct}\relax
\EndOfBibitem
\bibitem[Alml{\"o}f and Taylor(1991)Alml{\"o}f, and
  Taylor]{almlofAtomicNaturalOrbital1991}
Alml{\"o}f,~J.; Taylor,~P.~R. In \emph{Advances in {{Quantum Chemistry}}};
  L{\"o}wdin,~P.-O., Sabin,~J.~R., Zerner,~M.~C., Eds.; {Academic Press}, 1991;
  Vol.~22; pp 301--373\relax
\mciteBstWouldAddEndPuncttrue
\mciteSetBstMidEndSepPunct{\mcitedefaultmidpunct}
{\mcitedefaultendpunct}{\mcitedefaultseppunct}\relax
\EndOfBibitem
\bibitem[Richards \latin{et~al.}(1995)Richards, Kim, Yamaguchi, and
  Schaefer]{richardsDimethylcarbeneSingletGround1995}
Richards,~C.~A.; Kim,~S.-J.; Yamaguchi,~Y.; Schaefer,~H.~F. Dimethylcarbene:
  {{A Singlet Ground State}}? \emph{J. Am. Chem. Soc.} \textbf{1995},
  \emph{117}, 10104--10107\relax
\mciteBstWouldAddEndPuncttrue
\mciteSetBstMidEndSepPunct{\mcitedefaultmidpunct}
{\mcitedefaultendpunct}{\mcitedefaultseppunct}\relax
\EndOfBibitem
\bibitem[Lane(2013)]{laneCCSDTQOptimizedGeometry2013}
Lane,~J.~R. {{CCSDTQ Optimized Geometry}} of {{Water Dimer}}. \emph{J. Chem.
  Theory Comput.} \textbf{2013}, \emph{9}, 316--323\relax
\mciteBstWouldAddEndPuncttrue
\mciteSetBstMidEndSepPunct{\mcitedefaultmidpunct}
{\mcitedefaultendpunct}{\mcitedefaultseppunct}\relax
\EndOfBibitem
\bibitem[Miller(1975)]{millerSemiclassicalLimitQuantum1975}
Miller,~W.~H. Semiclassical Limit of Quantum Mechanical Transition State Theory
  for Nonseparable Systems. \emph{J. Chem. Phys.} \textbf{1975}, \emph{62},
  1899--1906\relax
\mciteBstWouldAddEndPuncttrue
\mciteSetBstMidEndSepPunct{\mcitedefaultmidpunct}
{\mcitedefaultendpunct}{\mcitedefaultseppunct}\relax
\EndOfBibitem
\bibitem[Miller(1998)]{millerQuantumSemiclassicalTheory1998}
Miller,~W.~H. Quantum and Semiclassical Theory of Chemical Reaction Rates.
  \emph{Faraday Discuss.} \textbf{1998}, \emph{110}, 1--21\relax
\mciteBstWouldAddEndPuncttrue
\mciteSetBstMidEndSepPunct{\mcitedefaultmidpunct}
{\mcitedefaultendpunct}{\mcitedefaultseppunct}\relax
\EndOfBibitem
\bibitem[Nguyen \latin{et~al.}(2010)Nguyen, Stanton, and
  Barker]{nguyenPracticalImplementationSemiclassical2010}
Nguyen,~T.~L.; Stanton,~J.~F.; Barker,~J.~R. A Practical Implementation of
  Semi-Classical Transition State Theory for Polyatomics. \emph{Chemical
  Physics Letters} \textbf{2010}, \emph{499}, 9--15\relax
\mciteBstWouldAddEndPuncttrue
\mciteSetBstMidEndSepPunct{\mcitedefaultmidpunct}
{\mcitedefaultendpunct}{\mcitedefaultseppunct}\relax
\EndOfBibitem
\end{mcitethebibliography}



\end{document}